\documentclass[aps,pra,reprint,groupedaddress, showpacs]{revtex4-1}
\usepackage{graphicx,graphics}
\usepackage{epstopdf} 
\usepackage{amsmath,amssymb,amsfonts}
\usepackage{latexsym,verbatim}
\usepackage{bm}
\usepackage{bbold}
\usepackage{color}
\usepackage[normalem]{ulem}
\usepackage[breaklinks=false,colorlinks,citecolor=blue,linkcolor=blue,urlcolor=blue]{hyperref}
\usepackage[abs]{overpic}
\usepackage{textcomp} 
\usepackage{mathtools}
\usepackage{braket}
\usepackage{soul}

\definecolor{Mygrey}{gray}{0.80}
\definecolor{lteal}{rgb}{0.10,0.60,0.70}
\definecolor{dkred}{rgb}{0.80,0.10,0.00}

 \ifdefined\LightComments

\else

 \fi

\begin{document}

\title{Spins extracted from fermionic states and their entanglement properties}

\author{Filippo Troiani}
\email{filippo.troiani@nano.cnr.it}
\author{Andrea Secchi}
\author{Stefano Pittalis}
\affiliation{Centro S3, CNR-Istituto di Nanoscienze, I-41125 Modena, Italy}

\begin{abstract} 
We investigate the spin states obtained by extracting $n$ electrons from closed-shell fermionic states. A partition of the system is defined through the introduction of the extraction modes. 
We derive the expression of the  $n$-body reduced density matrices, which represent the extracted spin states. 
We show that these states exhibit different forms of spin entanglement, whose detection is discussed in terms of the spin squeezing inequalities.
\end{abstract}

\date{\today}

\maketitle


\section{Introduction}\label{sec1}

Entanglement represents one of the resourceful hallmarks of quantum mechanics \cite{Nielsen_Chuang_2010,Horodecki09a}. The quantum-information oriented investigation of entanglement focuses on distinguishable sub-systems (qubits). On the other hand, the indistinguishability of identical particles makes the fundamental understanding and exploitation of entanglement in fermionic and bosonic systems more complex \cite{Benatti20a,PNASA2023}.
The case of fermions is most relevant for harnessing entanglement in quantum matter \cite{Schliemann01a, Li01a,Zanardi02a,Amico08a,Franca08a,Gigena15a,Franca11a}.
The so-called orbital entanglement \cite{Zanardi02a,Legeza2003,Rissler2006,Boguslawski2015} has been usefully applied to understand  bond formation processes \cite{Boguslawski2013}, chemical reactions \cite{Duperrouzel2015}, and active orbital selections \cite{Stein2016}. The approach has recently  been   revised to obtain precise estimations of the entanglements which may be  available a  resource for quantum technologies \cite{Ding2021}; an aspect which crucially relates to   super-selection rules, dictating the physically realizable states \cite{Wiseman03a,Bartlett03a,Friis2014,Friis2016}.
It was thus shown that a major part of the orbital entanglement seen previously may  not be available as resource \cite{Ding2021}.
 
Useful entanglement can result from the extraction of $n$ particles from an $N$-particle state \cite{Killoran14a,Bouvrie17a,Lo18a,Mahdavipour24a,Piccolini_2025,Troiani25a}. In this setting, the orbital degrees of freedom concur to determine the quantum properties of the extracted state but, more crucially, may provide labels. This allows one to regard the spins of the extracted particles as distinguishable objects, and apply to their state the entanglement theory developed for qubits. 
Identifying the labels with the position in space, the extraction of two fermions from a Fermi gas (2BRDM) results in rotationally-invariant two-spin states, characterized by an excess occupation of the singlet component. Provided that the two positions are sufficiently close to each other, such excess occupation results in a finite amount of two-spin entanglement \cite{Vedral03a,Shi04a}. This analysis was generalized to realistic inhomogenous systems, where much longer spin-entanglement lengths were determined in correspondence of atomic shells and molecular bonds \cite{Pittalis15a}. Further generalizations of this analysis to the case of arbitrary labels, defined by a generic set of orthonormal orbitals, and to more than two particles would be desirable, but are still missing.

In this work, we consider a generic closed-shell $N$-electron state, which can include the Fermi gas and atomic or molecular states as particular cases. Also, no restrictive assumptions are made on the modes that define the labels and the particle extraction: the present approach can thus be applied to different sets of physically- and experimentally-motivated orbitals \cite{Krylov20a}. Based on such partition, we first derive the form of the relevant reduced density operators (RDOs) corresponding to $n$-mode subsystems. The RDOs are constrained by the properties of the $N$-electron state, and specifically by its rotational invariance in the spin space, and by the defined values of the particle number and total spin projection. Thus we arrive at the expressions of the $n$-body reduced density matrices ($n$BRDMs), which define the relevant spin states.

The derived $n$BRDMs mirror the properties of the RDOs, and
can be written in a block diagonal form, each block being defined on the basis of the total spin, of its projection along the quantization axis, and of the mode occupations.  The entanglement properties of the resulting states display richer features, including --- for low values of the total spin and low particle numbers --- a high degree of genuine multipartite entanglement \cite{Troiani25a}. On the other hand, the weight of the high-spin components in the $n$BRDM increases with the number of doubly occupied modes in the RDOs. This, combined with the constraints resulting from the rotational invariance, tends to suppress entanglement, as shown through the use of spin-squeezing inequalities \cite{Guhne09a,Toth09a}.

The remainder of the paper is organized as follows.
In Sec. \ref{sec2}, we consider an $N$-electron state built from single or multiple configurations, each one consisting exclusively of doubly-occupied or unoccupied orbitals $\varphi_i$. Then a second set of orbitals $\phi_j$, related via a suitable unitary transformation to the orbitals $\varphi_i$, is identified with the individual subsystems (modes), which introduce an unambiguous partition of the fermionic system. 
Based on such partition, 
in Sec. \ref{sec3} we derive the reduced density operators for 
$n$-mode subsystems, with a detailed discussion of the cases up to $n=4$ (components of the RDOs that do not contribute to the relevant $n$BRDMs are disregarded).
In Sec. \ref{sec4}, we report the $n$RDMs
that can be related to the detection of $n$ particles localized at the orbitals $\phi_j$.  
Section \ref{sec5} discusses the entanglement properties of the detected spin states, and specifically the implications of rotational invariance and double occupations. Finally, Sec. \ref{sec6} reports the conclusions. Complementary information along with elementary technical details are provided in the Appendices \ref{appendixA}-\ref{appendixC}.

\section{From closed-shell to \\ more general singlet states} \label{sec2}

Let us consider a state of $N=2P$ electrons in a closed-shell state. Initially, we simplify our analysis by restricting it to a single Slater determinant, defined with respect to a set of $M \ge P$ orthogonal orbitals $\varphi_i({\bf r})$. Each of the first $P$ orbitals is occupied by two electrons, with opposite spin orientations, $\uparrow$ and $\downarrow$. 
In second quantization
\begin{align}\label{eqn:Psi}
    \big| \Psi \big> = \prod_{i = 1}^{P} \left( \hat{c}^\dagger_{i,\uparrow} \hat{c}^\dagger_{i,\downarrow} \right) \big| {\rm vac} \big> \,,
\end{align}
where $\hat{c}^\dagger_{i,\alpha}$ creates an electron in the orbital state $\varphi_i$ with spin $\alpha=\uparrow,\downarrow$. Note, $|\Psi\rangle$ thus corresponds to a product of singlet states, each one ``localized'' on an orbital $\varphi_i$. 
{\color{black} Closed-shell states in the form of a single Slater determinants, such as the one defined in the above equation, are at the base of widely used computational electronic structure methods, and allow one to characterize the ground states of a multitude of weakly correlated molecules and materials~\cite{Atkins,Martin}.
Besides, it was recently shown that - in spite of their apparent simplicity - closed-shell states consisting of a single Slater determinant allow the extraction of spin states that display maximal amounts of genuine multipartite entanglement \cite{Troiani25a}. 
}

\begin{figure}
\centering
\includegraphics[width=0.45\textwidth]{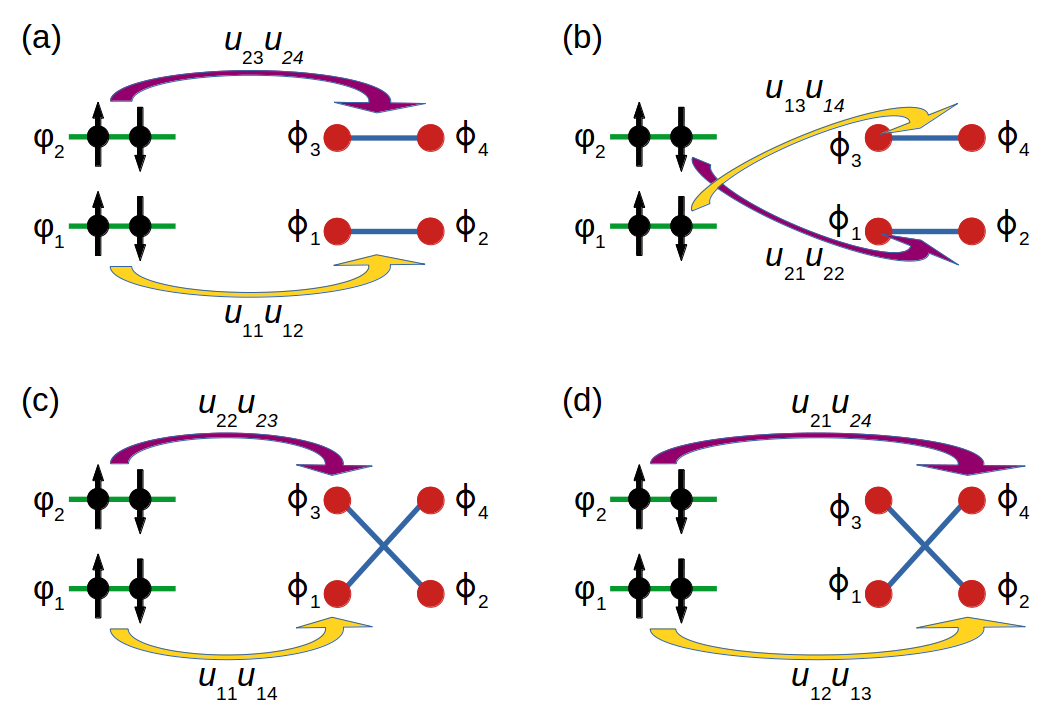}
\includegraphics[width=0.45\textwidth]{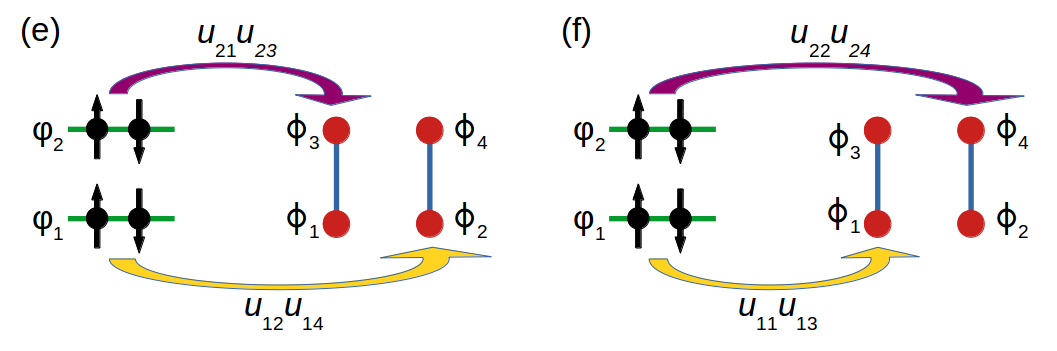}
\caption{The six kinds of contributions to the component of state $|\Psi\rangle$ corresponding to a single occupation of the orbitals $\phi_i$ ($M=N=4$): (a,b) $|\mathcal{S}_{12},\mathcal{S}_{34}\rangle$; (c,d) $|\mathcal{S}_{14},\mathcal{S}_{23}\rangle$; (e,f) $|\mathcal{S}_{13},\mathcal{S}_{24}\rangle$. The blue segments connecting pairs of orbitals $\phi_i$ denote the delocalized singlets. See the example discussed at the end of Sec. \ref{sec2}.}
\label{fig1}
\end{figure}

Now, let us consider a spin-independent change of the single-particle orbital basis, from $\{\varphi_i\}$ to another $M$-dimensional orthonormal basis $\{\phi_j\}$, defined by a unitary matrix $U$. {\color{black} The orbitals $\phi_i$ define the extraction modes, whose spin state is given by the $n$BRDM. These orbitals, whose physical properties need not be specified within the present analysis, act as labels for $n$ extracted electrons, thus allowing the mapping from a system of indistinguishable particles (fermions) to one of distinguishable particles (spins). From a physical point of view, the identification of the $\phi_i$ with a well defined set of localized orbitals, within atomistic or model systems, provides a more concrete meaning to the spin entanglement of the extracted particles \cite{Troiani25a}. From a formal point of view,} we stress that the transformation $U$ involves not only the occupied orbitals but also all the unoccupied ones. Correspondingly,  the creation operators transform as follows:
\begin{align}\label{eq:chbasis}
\hat{c}_{i,\alpha}^\dagger = \sum_{j=1}^M u_{ij}\, \hat{d}_{j,\alpha}^\dagger\,.
\end{align}
It is worth noticing that
\begin{align}
    \hat{c}^\dagger_{i,\uparrow}  \hat{c}^\dagger_{i,\downarrow} &=  \sum_{j,k = 1}^{M} u_{ij} u_{ik} \hat{d}^{\dagger}_{j, \uparrow}\, \hat{d}^{\dagger}_{k, \downarrow} 
=   \sum_{j,k = 1}^{M} u_{ij} u_{ik}\, \hat{S}^{\dagger}_{j,k}  \,,
\end{align}
where we have introduced the {\it singlet creation operators} (SCOs), defined by
\begin{align}
    \hat{S}^{\dagger}_{j,k} \equiv \frac{1}{2} \left( \hat{d}^{\dagger}_{j, \uparrow} \hat{d}^{\dagger}_{k, \downarrow} - \hat{d}^{\dagger}_{j, \downarrow} \hat{d}^{\dagger}_{k, \uparrow}  \right)    \,.
    \label{SCO}
\end{align}
These create a localized spin singlet if $j = k$, or a ``delocalized'' spin singlet if $j < k$:
\begin{align}
    |\mathcal{S}_{jj}\rangle\equiv \hat{S}^{\dagger}_{j,j} |{\rm vac}\rangle\,,\ \ \  |\mathcal{S}_{jk}\rangle\equiv \sqrt{2}\,\hat{S}^{\dagger}_{j,k} |{\rm vac}\rangle =|\mathcal{S}_{kj}\rangle\,.
\end{align}
The state $|\Psi\rangle$  in Eq.~\eqref{eqn:Psi} can thus be expressed as \cite{Troiani25a}
\begin{align}
    \big| \Psi \big> = \left( \prod_{i=1}^{P} \sum_{j_i=1}^{M} \sum_{k_i=1}^{M} u_{ij_i} u_{ik_i}  \hat{S}^{\dagger}_{j_i, k_i} \right) \big| {\rm vac} \big>  \,.
    \label{Psi in terms of d}
\end{align}
This is a linear superposition of different configurations, each one given by the product of $P$ localized or delocalized singlets.

To dig into Eq.~\eqref{Psi in terms of d},  note that the SCOs are invariant with respect to exchange of the indices and {\em commute} with each other
\begin{align}
    & \hat{S}^{\dagger}_{j,k} = \hat{S}^{\dagger}_{k,j} \,, \quad
    \hat{S}^{\dagger}_{j_1,k_1} \hat{S}^{\dagger}_{j_2,k_2} = \hat{S}^{\dagger}_{j_2,k_2} \hat{S}^{\dagger}_{j_1,k_1}   \,.
    \label{properties singlet operators}
\end{align}
Besides, the product of two SCOs can be written as
\begin{align}
    & \hat{S}^{\dagger}_{j_1, k_1} \hat{S}^{\dagger}_{j_2,k_2} = - \frac{1}{4} \left( 1 + P_{\uparrow \downarrow} \right) \times \nonumber \\ & \quad \Big(   \hat{d}^{\dagger}_{j_1, \uparrow} \hat{d}^{\dagger}_{j_2, \uparrow} \hat{d}^{\dagger}_{k_1, \downarrow} \hat{d}^{\dagger}_{k_2, \downarrow} 
    +   \hat{d}^{\dagger}_{j_1, \uparrow} \hat{d}^{\dagger}_{k_2, \uparrow} \hat{d}^{\dagger}_{k_1, \downarrow} \hat{d}^{\dagger}_{j_2, \downarrow}  \Big) \,,
    \label{product two singlet operators}
\end{align} 
where the symbol $P_{\uparrow \downarrow}$ permutes the spin indices $\uparrow$ and $\downarrow$ in the operators appearing on its right hand side.
Equation \eqref{product two singlet operators}  allows us to verify some remarkable identities. 
In fact, the product $\hat{S}^{\dagger}_{j_1, k_1} \hat{S}^{\dagger}_{j_2,k_2}$ vanishes if and only if three or four of the indices $(j_1,k_1,j_2,k_2)$ coincide. If there are two (or two pairs of) equal indices, then it is possible to rewrite the above product so as to introduce SCOs that create local singlets, being 
\begin{align}
    \hat{S}^{\dagger}_{j, k_1} \hat{S}^{\dagger}_{j,k_2} = {\color{black}-} \frac{1}{2} \hat{S}^{\dagger}_{j, j} \hat{S}^{\dagger}_{k_1,k_2} \,.
    \label{remarkable}
\end{align}
The above equation holds for arbitrary values of $k_1$ and $k_2$, provided that they both differ from $j$.
A spin singlet localized in the mode $\phi_j$ can thus result either directly from one of the doubly occupied orbitals $\varphi_i$ (with amplitude {\color{black} proportional to} $u_{ij}^2$), or from the creation of partially overlapping delocalized singlets related to different orbitals $\varphi_i$ and $\varphi_l$ (with amplitude {\color{black} proportional to} $u_{ij} u_{lj}$). 
For example, a product of four SCOs with partially overlapping indices can be transformed as follows into the product of disjoint SCOs:
\begin{align}
    \hat{S}^{\dagger}_{j_1,k_1} \hat{S}^{\dagger}_{j_1,k_2} \hat{S}^{\dagger}_{j_3,k_3}  \hat{S}^{\dagger}_{k_3,j_3} = \frac{1}{4} \hat{S}^{\dagger}_{j_1,j_1} \hat{S}^{\dagger}_{j_3,j_3}  \hat{S}^{\dagger}_{k_3,k_3}\hat{S}^{\dagger}_{k_1,k_2}  \,.
\end{align}
This relation, derived from Eqs.~\eqref{properties singlet operators} and \eqref{remarkable}, applies if $j_1$, $j_3$, and $k_3$ differ from each other and from all the remaining indices. If this is not the case, the above product vanishes. 

From the above property it follows that any term $|\Phi\rangle \equiv \prod_{i = 1}^{P}  \hat{S}^{\dagger}_{j_i, k_i} \big| {\rm vac} \big> $  in Eq.~\eqref{Psi in terms of d}: $(i)$ vanishes, if any index appears more than twice in the product; $(ii)$ can be rearranged, by repeated application of Eq.~\eqref{remarkable}, in an expression of the form
\begin{align} 
    |\Phi\rangle = \frac{\color{black}(-1)^p}{2^p} \,\hat{S}^{\dagger}_{j_1',j_1'} \ldots \hat{S}^{\dagger}_{j_p',j_p'} 
    \hat{S}^{\dagger}_{j_{p+1}',k_{p+1}'} \ldots \hat{S}^{\dagger}_{j_P',k_P'}\,\big| {\rm vac} \big> \quad \nonumber \\
    \equiv \frac{1}{2^p} \frac{1}{2^{q/2}} | \mathcal{S}_{j_1'\,j_1'} \ldots \mathcal{S}_{j_p'\,j_p'} 
    \mathcal{S}_{j_{p+1}'\,k_{p+1}'} \ldots \mathcal{S}_{j_P'\,k_P'} \rangle \,,
    \label{eq:gss}
\end{align}
{\color{black} being $p$ the number of indices that appear twice among the pairs $(j_i,k_i)$, for two different values of $i$. 
The above expression} is a product of $p$ localized (single-mode) singlets $\mathcal{S}_{j_n j_n}$ with $n \in \lbrace 1, \ldots, p \rbrace$, and $q=P-p$ delocalized (two-mode) singlets $\mathcal{S}_{j_n k_n}$ with $n \in \lbrace p+1, \ldots, P \rbrace$, where all the indices $j_n$ and $k_n$ differ from each other.

We conclude that, upon a spin-independent unitary transformation of the orbital basis [Eq.~\eqref{eq:chbasis}], a single-configurational product of localized singlets [Eq.~\eqref{eqn:Psi}]  transforms into a multi-configurational state, whose components correspond to the products of localized and delocalized singlets [Eq.~\eqref{eq:gss}].
The same conclusion applies to 
{\color{black} any} multi-configurational state {\color{black} consisting of a linear superposition of} Slater determinants{\color{black}, each one} involving only doubly-occupied orbitals.

\paragraph*{Example: the four-electron case.}
In order to illustrate the above equations, we consider the case where $N=M=4$ and the unitary transformation coincides with the quantum Fourier transform: $ u_{jk} = \frac{1}{2}\,e^{i\pi (j-1)(k-1)/2}$. The expression of the state $|\Psi\rangle$ given in Eq. (\ref{Psi in terms of d}) includes $N^M=256$ contributions, 112 of which vanish identically because of they would imply the double occupation of at least one spin-orbital. Focusing on the 24 terms that correspond to a single occupation of all four orbitals $\phi_i$ (Fig. \ref{fig1}), one obtains:
\begin{align}
    |\Psi\rangle = \frac{i}{8} (|\mathcal{S}_{12},\mathcal{S}_{34}\rangle - |\mathcal{S}_{14},\mathcal{S}_{23}\rangle ) + \dots\,.
\end{align}
The additional contributions, which are not explicitly reported in the above equation, correspond to components that include one or two localized singlets, and an equal number of unoccupied orbitals $\phi_i$. 

{

\section{Reduced density operators\label{sec3}}

To initiate our analysis of the entanglement properties,
the system of orbitals is partitioned into subsystems, each of which coincides with one or more orbitals (or modes) $\phi_k$. If the latter are localized in space, one may identify the subsystem with the sites of a lattice.
Each mode $k$ can thus be found in one of four different states $|n_{k,\uparrow},n_{k,\downarrow}\rangle$, characterized by the spin-resolved occupation numbers $n_{k,\uparrow}=0,1$ and $n_{k,\downarrow}=0,1$: $|u_k\rangle\equiv|0,0\rangle$ (unoccupied orbital), $|\!\Uparrow_k\rangle\equiv |1,0\rangle$ (orbital occupied by one, spin-up electron), $|\!\Downarrow_k\rangle\equiv|0,1\rangle$ (orbital occupied by one, spin-down electron), $|\mathcal{S}_{kk}\rangle\equiv |1,1\rangle$ (orbital occupied by two electrons, forming a spin singlet). 

Correspondingly, we may consider the projection operators:
\begin{align}\label{pr1}
    \hat{\mathcal{P}}_{k|2} = |\mathcal{S}_{kk}\rangle\langle \mathcal{S}_{kk}| \,,\ 
    \hat{\mathcal{P}}_{k|\Uparrow} = |\!\Uparrow_{k}\rangle\langle \Uparrow_{k}\!| \,, \\
    \hat{\mathcal{P}}_{k|\Downarrow} = |\!\Downarrow_{k}\rangle\langle \Downarrow_{k}\!| \,,\  
    \hat{\mathcal{P}}_{k|0} = |u_{k}\rangle\langle u_{k}|\,.
    \label{pr2}
\end{align}
In the expressions of the reduced density operators, we shall also make use of the operators
$\hat{\mathcal{P}}_{ij \dots k|S,M}$, defined as the projectors on the {\color{black}totally symmetric} state with $S=m/2$ and $-S\le M\le S$, formed by $m$ spins, each one localized in one of the $m$ orbitals $(\phi_i,\phi_j \dots \phi_k)$.

\subsection{General properties}

\begin{figure}
\centering
\includegraphics[width=0.4\textwidth]{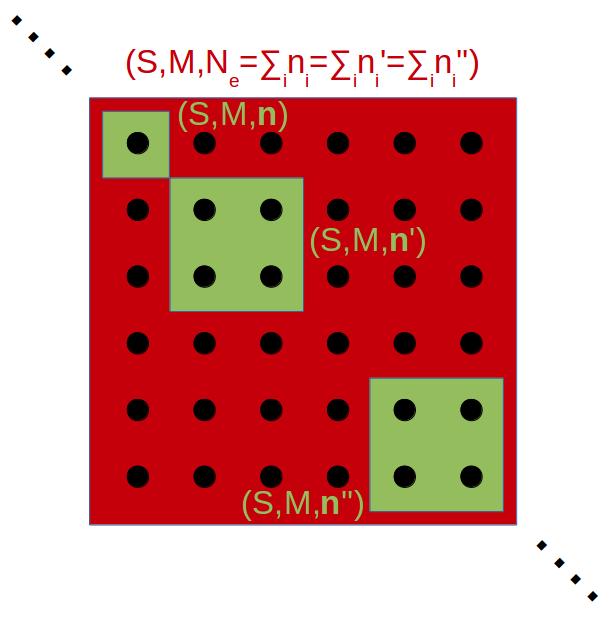}
\caption{Block-diagonal form of the reduced density matrices $\rho_{ijk\dots}$ derived from closed-shell states $|\Psi\rangle$. Nonzero elements can only be present within subspaces defined by the values of the total spin $S$, its projection $M$ along the quantization axis, and the particle number $N_e$ (red area). The elements that contribute to the $n$BRDM $\hat\Gamma_{ijk\dots}$ are a subset of these nonzero elements, and are all contained in subspaces with identical mode occupation vectors ${\bf n}$, characterized by all nonzero components ($n_l\neq 0$ for $l=i,j,\dots,k$, green areas).}
\label{fig2}
\end{figure}

The reduced density operators (RDOs) that define the state of $K$ modes are obtained from the density operator $\hat{\rho} = |\Psi\rangle\langle\Psi|$ ($\Psi$ being an $N$-particle state) by performing a partial trace over the complementary modes. 

The RDOs is block diagonal, because some kind of off-diagonal terms vanish identically. As shown in Appendix \ref{appendixA}, a RDO can only include off-diagonal elements (also referred to as {\it coherences}) between pairs of states characterized by identical values of the total spin projection ($M$), of the particle number ($N_e$), and of the total spin ($S$). The first two constraints follow from the fact that the total spin projection and the particle number have well defined values in the density operator 
$\hat{\rho} = |\Psi\rangle\langle\Psi|$; 
the latter constraint results from the rotational invariance in the spin space of the state $\hat{\rho}$ and (thus) of the RDOs.

There are off-diagonal terms in the RDOs that are generally nonzero, but do not contribute to the $n$BRDMs. These are the ones that involve pairs of states with equal particle numbers but different occupation of (at least one of) the modes. Being the derivation of the $n$BRDMs (Sec. \ref{sec4}) the main objective of the present investigation, we disregard such off-diagonal terms in the discussion of the RDOs. These can thus be divided in blocks (subspaces), each one defined by a given $\zeta=(S,M,{\bf n})$, being ${\bf n} = (n_i,n_j,\dots)$ a vector, whose components represent the occupation numbers of the relevant modes (Fig. \ref{fig2}). We note in passing that these RDOs fully account for the parity and particle-number superselection rules~\cite{Bartlett03a,Friis2014,Friis2016,Ding2021}. {\color{black} Besides, the identification of a subsystem with an orbital mode $k$, each one corresponding to the two spin-orbitals $(k,\uparrow)$ and $(k,\downarrow)$, combined with the absence of coherences between subspaces defined by different $\zeta$, removes the ambiguity related to the anticommutation properties of the fermionic creation and annihilation operators~\cite{Friis13a,Szalay_2021}.}

In the following Subsections,  for illustrative purposes, we start by discussing the approach and the notation in the cases of one and two-mode subsystems \cite{Rissler2006,Boguslawski2015}; then --- toward the analyses of entanglement --- we consider the three- and four-mode cases; we finally generalize to arbitrary number of subsystems. 
In all cases, we disregard the components of the RDOs that include unoccupied modes, which do not contribute to the $n$BRDMs. Therefore, for the $n$-mode subsystem, we only need to consider the particle numbers ranging from $n$ to $2n$. We refer to the RDOs, without the off-diagonal terms between different subspaces $\zeta$ and without unoccupied-mode components, with the symbol $\hat{\rho}_{ij\dots}$.

\subsection{Single-mode subsystems\label{subsecA}}

The reduced density operator of a single mode $i$ (orbital $\phi_i$) is in general given by a mixture of states corresponding to up to two electrons:
\begin{align}
    \hat{\rho}_i 
= \sum_\zeta p_\zeta^{(1)} \,\hat{\rho}_\zeta^{(1)} \,,
\end{align}
where $\zeta = (S,M;n_i)$ specifies the values of the total spin, of its projection along the quantization axis, and the mode occupation; $p^{(1)}_{\zeta}$ and $\hat{\rho}^{(1)}_{\zeta}$ are, respectively, the probability and the density matrix associated to subspace $\zeta$. The subspace density matrices are listed in the following.

\subsubsection{One electron}
\paragraph{\color{black} Subspace $S=1/2$.}
The single-particle contribution defined within the subspace $\zeta = (1/2,M;1) $ is given by:
\begin{align}    
\hat{\rho}^{(1)}_{1/2,M;1} = 
\hat{\mathcal{P}}_{i|1/2,M} \,,
\end{align}
where $\hat{\mathcal{P}}_{i|1/2,1/2}=\hat{\mathcal{P}}_{i|\Uparrow}$ and $\hat{\mathcal{P}}_{i|1/2,-1/2}=\hat{\mathcal{P}}_{i|\Downarrow}$. Due to rotational invariance, 
$p_{1/2,1/2;1} = p_{1/2,-1/2;1}$. 

\subsubsection{Two electrons}
\paragraph{\color{black} Subspace $S=0$.}
The two-electron term can be identified with the projector on the singlet state localized in the mode $i$, and thus with $\zeta = (0,0;2)  $: 
\begin{align}
\hat{\rho}_{0,0;2}^{(1)} = 
\hat{\mathcal{P}}_{i|2}  .
\end{align}

\subsection{Two-mode subsystem\label{subsecB}}

The reduced density operator of two modes $i$ and $j$ (orbitals $\phi_i$ and $\phi_j$), is in general given by a mixture of states corresponding to up to four electrons, each one including different contributions:
\begin{align}
\hat{\rho}_{ij} 
= \sum_\zeta p_{\zeta}^{(2)} \hat{\rho}_{\zeta}^{(2)} \,,
\end{align}
where $\zeta=(S,M;n_i,n_j)$ specifies the values of the total spin, of its projection along the quantization axis, and the mode occupations.  

\subsubsection{Two electrons}
\paragraph{\color{black} Subspace $S=0$.}
The two-electron singlet component of the RDO is defined within the subspace generated by the states $|\mathcal{S}_{ii}\rangle$, $|\mathcal{S}_{ij}\rangle$, and $|\mathcal{S}_{jj}\rangle$. Disregarding the states that include unoccupied modes, one can identify such contribution with the term:
\begin{align}
\hat{\rho}_{0,0;1,1}^{(2)} 
= |\mathcal{S}_{ij} \rangle\langle \mathcal{S}_{ij} | \,.
\end{align}

\paragraph{\color{black} Subspace $S=1$.}
The triplet component can be written as a combination of three terms, each one corresponding to a different value of the total-spin projection $M$:
\begin{align}
\hat{\rho}_{1,M;1,1}^{(2)} = \hat{\mathcal{P}}_{ij|1,M}\,,
\end{align}
where, for example, $\hat{\mathcal{P}}_{ij|1,1} = \hat{\mathcal{P}}_{i|\Uparrow}\, \hat{\mathcal{P}}_{j|\Uparrow}$. Due to rotational invariance, the three probabilities $p_{1,M;1,1}$, with $M=0,\pm 1$, must coincide. 

\subsubsection{Three electrons}
\paragraph{\color{black} Subspace $S=1/2$.}
The three-electron contribution is defined within the two-dimensional $S=1/2$ subspace spanned by the states $|\alpha_i,\mathcal{S}_{jj}\rangle$ and $|\mathcal{S}_{ii},\alpha_j\rangle$ (where $\alpha = \Uparrow,\Downarrow$ and $\alpha_i=\alpha_j$). The contribution of the RDO within this subspace can be written as a combination of two terms that differ in terms of mode occupation. For ${\bf n}=(2,1)$, one has:
\begin{align}
\hat{\rho}_{1/2,M;2,1}^{(2)} = 
\hat{\mathcal{P}}_{i|2}\,\hat{\mathcal{P}}_{j|1/2,M}\,.
\end{align}
The contributions corresponding to ${\bf n}=(1,2)$ are obtained simply by exchanging the indices $i$ and $j$ in the above expression. Due to rotational invariance, the probabilities related to subspaces that differ only in $M$ must coincide.

\subsubsection{Four electrons}
\paragraph{\color{black} Subspace $S=0$.}
The four-electron contribution corresponds to both modes being in the singlet state:
\begin{align}
    \hat{\rho}_{0,0;2,2}^{(2)} = 
    \hat{\mathcal{P}}_{i|2}\,\hat{\mathcal{P}}_{j|2}.
\end{align}

\subsection{Three-mode subsystems\label{subsecC}}

\begin{table}
\centering
\begin{tabular}{|c|cc|c|}
\hline
\multicolumn{4}{|c|}{${\bf n}=(1,1,1)\ \ \ \ \ [z(3,3)=1]$ }\\
\hline
Subspace \# & $S$ & $d_S$ & State $M\!=\!S$ (example) \\
\hline
1 & $1/2$ & 2 & $|\mathcal{S}_{ij},\Uparrow_k\rangle$ \\
2 & $3/2$ & 1 & $|\Uparrow_i,\Uparrow_j,\Uparrow_k\rangle$ \\
\hline
\hline
\multicolumn{4}{|c|}{${\bf n}=(2,1,1)\ \ \ \ \ [z(3,4)=3]$ }\\
\hline
Subspace \# & $S$ & $d_S$ & State $M\!=\!S$ (example) \\
\hline
3-5 & $0$ & 1 & $|\mathcal{S}_{ii},\mathcal{S}_{jk}\rangle$ \\
6-8 & $1$ & 1 & $|\mathcal{S}_{ii},\Uparrow_j,\Uparrow_k\rangle$ \\
\hline
\hline
\multicolumn{4}{|c|}{${\bf n}=(2,2,1)\ \ \ \ \ [z(3,5)=3]$ }\\
\hline
Subspace \# & $S$ & $d_S$ & State $M\!=\!S$ (example) \\
\hline
9-11 & $1/2$ & 1 & $|\mathcal{S}_{ii},\mathcal{S}_{jj},\Uparrow_k\rangle$ \\
\hline
\hline
\multicolumn{4}{|c|}{${\bf n}=(2,2,2)\ \ \ \ \ [z(3,6)=1]$ }\\
\hline
Subspace \# & $S$ & $d_S$ & State $M\!=\!S$ (example) \\
\hline
12 & $0$ & 1 & $|\mathcal{S}_{ii},\mathcal{S}_{jj},\mathcal{S}_{kk}\rangle$ \\
\hline
\end{tabular}
\caption{\label{TableI} Subspaces corresponding to three modes (orbitals $\phi_i$, $\phi_j$, and $\phi_k$). Each subspace is defined by the total spin $S$, its projection $M=S$, and by the site occupation numbers ${\bf n}$. The numbers $d_S$, and $z(n,N_e)$ are the multiplicity related to scalar quantum numbers (partial spin sums) and the number of possible mode occupations ${\bf n}$, respectively. The overall numbers of subspaces that block diagonalize the RDO and the 3BRDM, including the multiplicity {\color{black}$(2S+1)$} related to the total spin projection, are $A=25$ and $B=6$, respectively.}
\end{table}

The reduced density operator for the three modes ($i$, $j$, and $k$) can be written as a mixture of different terms, corresponding to up to six electrons:
\begin{align}
\hat{\rho}_{ijk} 
= \sum_{\zeta} p_{\zeta}^{(3)} \hat{\rho}_{\zeta}^{(3)} \,,
\end{align}
where the subscript $\zeta=(S,M;n_i,n_j,n_k)$ specifies the overall spin $S$, its projection $M$, and the site occupations. The complete list of subsystems $\zeta$ that block diagonalize $\hat{\rho}_{ijk}$ is given in Table \ref{TableI}.

\subsubsection{Three electrons}
\paragraph{\color{black} Subspace $S=1/2$.}
The three-electron terms that include one singlet state are defined in the two-dimensional subspaces $\zeta = (1/2,\pm 1/2;1,1,1)$, spanned by the non-orthogonal states $\{|\mathcal{S}_{ij},\alpha_k\rangle,|\alpha_i,\mathcal{S}_{jk}\rangle,|\mathcal{S}_{ik},\alpha_j\rangle\}$,
with $\alpha=\Uparrow,\Downarrow$ and $\alpha_i=\alpha_j=\alpha_l$. 

\paragraph{\color{black} Subspace $S=3/2$.}
The three-electron terms that include no singlet states is defined within the subspace{\color{black}s}  $\zeta = (3/2,M;1,1,1)$. 
The generic density matrix within {\color{black}each of these} subspace{\color{black}s} can be written as:
\begin{align}
\hat{\rho}_{3/2,M;1,1,1}^{(3)} = \hat{\mathcal{P}}_{ijk|3/2,M} \,,
\end{align}
where, for example, $\hat{\mathcal{P}}_{ijk|3/2,3/2} = \hat{\mathcal{P}}_{i|\Uparrow} \,\hat{\mathcal{P}}_{j|\Uparrow}\,\hat{\mathcal{P}}_{k|\Uparrow}$. Due to rotational invariance, the four probabilities $p_{3/2,M;1,1,1}$ must all be identical.

\subsubsection{Four electrons}
\paragraph{\color{black} Subspace $S=0$.}
The four-electron term that includes two singlet states {\color{black}(a localized and a delocalized one)} is defined in the subspace $S=0$, spanned by $|\mathcal{S}_{ii},\mathcal{S}_{jk}\rangle$, $|\mathcal{S}_{ik},\mathcal{S}_{jj}\rangle$, $|\mathcal{S}_{ij},\mathcal{S}_{kk}\rangle$. 
Disregarding the coherences between states characterized by different occupations of the three modes, the density operator within this subspace can be written as a combination of the three projectors, one for each ${\bf n}$. In particular, for $\zeta=(0,0;2,1,1)$ one has: 
\begin{align}
\hat{\rho}_{0,0;2,1,1}^{(3)} = |\mathcal{S}_{jk}\rangle\langle \mathcal{S}_{jk}| \otimes \hat{\mathcal{P}}_{i|2}\,.
\end{align}

\paragraph{\color{black} Subspace $S=1$.}
The four-electron term that includes only one {\color{black}(localized)} singlet state is defined in the $S=1$ subspace spanned by {\color{black} the symmetrized components of} $|\mathcal{S}_{ii},\alpha_j,\alpha_k\rangle$, $|\alpha_i,\mathcal{S}_{jj},\alpha_k\rangle$, {\color{black} and} $|\alpha_i,\alpha_j,\mathcal{S}_{kk}\rangle$. For example, the RDO within the subspace $\zeta=(1,M;2,1,1)$ can be written as:
\begin{align}
\hat{\rho}_{1,M;2,1,1}^{(3)} = \hat{\mathcal{P}}_{i|2}\,\hat{\mathcal{P}}_{jk|1,M}\,.
\end{align}
Analogous terms are obtained for ${\bf n} = (1,2,1)$ and ${\bf n} = (1,1,2)$, simply by exchanging the index $i$ with either $j$ or $k$.
Considering the 3 possible values of $M$ and the 3 {\color{black}the three vectors} ${\bf n}$, the $S=1$ and $N_e=4$ sector includes 9 different subspaces.

\subsubsection{Five electrons}
\paragraph{\color{black} Subspace $S=1/2$.}
The five-electron includes two localized singlet states and is defined in the $S=1/2$ subspaces, spanned by the states 
$|\mathcal{S}_{ii},\mathcal{S}_{jj},\alpha_k\rangle$, $|\mathcal{S}_{ii},\alpha_j,\mathcal{S}_{kk}\rangle$, and
$|\alpha_i,\mathcal{S}_{jj},\mathcal{S}_{kk}\rangle\}$, 
with $\alpha_i=\alpha_j=\alpha_k=\Uparrow,\Downarrow$.

For example, the RDOs within the subspace ${\bf n}=(2,2,1)$ can be written as:
\begin{align}
\hat{\rho}_{1/2,M;2,2,1}^{(3)}=
\hat{\mathcal{P}}_{i|2}\,\hat{\mathcal{P}}_{j|2}\,\hat{\mathcal{P}}_{k|1/2,M} \,.
\end{align}
Analogous terms are obtained for ${\bf n} = (2,1,2)$ and ${\bf n} = (1,2,2)$, simply by exchanging the index $k$ with either $i$ or $j$.
Considering the 2 possible values of $M$ and the three {\color{black} vectors} ${\bf n}$, the $S=1$ and $N_e=4$ sector includes 6 different subspaces.

\subsubsection{Six electrons}
\paragraph{\color{black} Subspace $S=0$.}
The six-electron term belongs to the singlet subspace, and is specifically given by:
\begin{align}
    \hat{\rho}_{0,0;2,2,2}^{(3)} =  \hat{\mathcal{P}}_{i|2}\,\hat{\mathcal{P}}_{j|2}\,\hat{\mathcal{P}}_{k|2}\,,
\end{align}
which corresponds to the projector on the three localized singlet states. 

\subsection{Four-mode subsystems\label{subsecD}}

\begin{table}[]
\centering
\begin{tabular}{|c|cc|c|}
\hline
\multicolumn{4}{|c|}{${\bf n}=(1,1,1,1)\ \ \ \ \ [z(4,4)=1]$ }\\
\hline
Subspace \# & $S$ & $d_S$ & State $M\!=\!S$ (example) \\
\hline
1 & $0$ & 2 & $|\mathcal{S}_{ij},\mathcal{S}_{kl}\rangle$ \\
2 & $1$ & 3 & $|\mathcal{S}_{ij},\Uparrow_k,\Uparrow_l\rangle$ \\
3 & $2$ & 1 & $|\Uparrow_i,\Uparrow_j,\Uparrow_k,\Uparrow_l\rangle$ \\
\hline
\hline
\multicolumn{4}{|c|}{${\bf n}=(2,1,1,1)\ \ \ \ \ [z(4,5)=4]$ }\\
\hline
Subspace \# & $S$ & $d_S$ & State $M\!=\!S$ (example) \\
\hline
4-7 & $1/2$ & 2 & $|\mathcal{S}_{ii},\mathcal{S}_{jk},\Uparrow_l\rangle$ \\
8-11 & $3/2$ & 1 & $|\mathcal{S}_{ii},\Uparrow_j,\Uparrow_k,\Uparrow_l\rangle$ \\
\hline
\hline
\multicolumn{4}{|c|}{${\bf n}=(2,2,1,1)\ \ \ \ \ [z(4,6)=6]$ }\\
\hline
Subspace \# & $S$ & $d_S$ & State $M\!=\!S$ (example) \\
\hline
12-17 & $0$ & 1 & $|\mathcal{S}_{ii},\mathcal{S}_{jj},\mathcal{S}_{kl}\rangle$ \\
18-23 & $1$ & 1 & $|\mathcal{S}_{ii},\mathcal{S}_{jj},\Uparrow_k,\Uparrow_l\rangle$ \\
\hline
\hline
\multicolumn{4}{|c|}{${\bf n}=(2,2,2,1)\ \ \ \ \ [z(4,7)=4]$ }\\
\hline
Subspace \# & $S$ & $d_S$ & State $M\!=\!S$ (example) \\
\hline
24-27 & $1/2$ & 1 & $|\mathcal{S}_{ii},\mathcal{S}_{jj},\mathcal{S}_{kk},\Uparrow_l\rangle$ \\
\hline
\hline
\multicolumn{4}{|c|}{${\bf n}=(2,2,2,2)\ \ \ \ \ [z(4,8)=1]$ }\\
\hline
Subspace \# & $S$ & $d_S$ & State $M\!=\!S$ (example) \\
\hline
28 & $0$ & 1 & $|\mathcal{S}_{ii},\mathcal{S}_{jj},\mathcal{S}_{kk},\mathcal{S}_{ll}\rangle$ \\
\hline
\end{tabular}
\caption{\label{TableII} Subspaces corresponding to four modes (orbitals $\phi_i$, $\phi_j$, $\phi_k$, and $\phi_l$). Each subspace is defined by the total spin $S$, its projection $M=S$, and by the site occupation numbers ${\bf n}$. The numbers $d_S$ and $z(n,N_e)$ are the multiplicity related to scalar quantum numbers (partial spin sums) and the number of possible mode occupations ${\bf n}$. The overall numbers of subspaces that block diagonalize the RDO and the 4BRDM, including the multiplicity {\color{black}$(2S+1)$} related to the total spin projection, are $A=66$ and $B=9$, respectively.}
\end{table}

The reduced density operator for the four modes ($i$, $j$, $k$, and $l$) is in general given by the mixture of different components, corresponding to up to eight electrons:
\begin{align}
    \hat{\rho}_{ijkl} 
= \sum_{\zeta} p_{\zeta}^{(4)} \hat{\rho}_{\zeta}^{(4)} \,,
\end{align}
where the subscript $\zeta=(S,M;n_i,n_j,n_k,n_l)$ specifies the overall spin $S$, its projection $M$, and the site occupation. The full list of subsystems that block diagonalize $\hat{\rho}_{ijkl}$ is given in Table \ref{TableII}.

\subsubsection{Four electrons}
\paragraph{\color{black} Subspace $S=0$.}
The four-electron terms that include two singlet states are defined in the two-dimensional subspace $\zeta=(0,0;1,1,1,1)$, spanned by the non-orthogonal states $\{|\mathcal{S}_{ij},\mathcal{S}_{kl}\rangle, |\mathcal{S}_{ik},\mathcal{S}_{jl}\rangle, |\mathcal{S}_{il},\mathcal{S}_{kj}\rangle\}$. 

\paragraph{\color{black} Subspace $S=1$.}
The four-electron terms that include only one singlet state are defined in the three-dimensional subspaces $\zeta=(1,M;1,1,1,1)$, spanned by the orthogonal states like $|\mathcal{S}_{ij},\mathcal{S}_{kl}\!=\!1,M_{kl}\!=\!M\rangle$ or $|\mathcal{S}_{jk},\mathcal{S}_{il}\!=\!1,M_{il}\!=\!M\rangle$, where $M_{ij}$ ($M_{il}$) is the projection of the partial spin sum $\mathcal{S}_{ij}$ ($\mathcal{S}_{il}$). 

\paragraph{\color{black} Subspace $S=2$.}
The four-electron terms that include no singlet states are defined in the $\zeta=(2,M;1,1,1,1)$ subspace. 
The generic rotationally-invariant density operator within this subspace can be written as a combination of the terms:
\begin{align}
\hat{\rho}^{(4)}_{2,M;1,1,1,1} = \hat{\mathcal{P}}_{ijkl|2,M}\,.
\end{align}
Due to rotational invariance, all the {\color{black}probabilities} $p_{2,M;1,1,1,1}$ (corresponding to different values of $M$) coincide.

\subsubsection{Five electrons}
\paragraph{\color{black} Subspace $S=1/2$.}
The five-electron terms that include two singlets, a localized and a delocalized one, are defined in subspaces such as $\zeta=(1/2,M;2,1,1,1)$, which are spanned by the three states $|\mathcal{S}_{ii},\mathcal{S}_{ab},\alpha_c\rangle$ ($c=j,k,l$ and $\alpha_i=\alpha_j=\alpha_k$). 

\paragraph{\color{black} Subspace $S=3/2$.}
The five-electron terms with only one {\color{black}localized} singlet state are defined in the subspaces such as $\zeta=(3/2,M;2,1,1,1)$. 
The generic density operator within this subspace can be written as:  
\begin{align}
    \hat{\rho}_{3/2,M;2,1,1,1}^{(4)} =  \hat{\mathcal{P}}_{i|2}\,\hat{\mathcal{P}}_{jkl|3/2,M}\,.
\end{align}
The rotational invariance of the state $|\Psi\rangle$ in the spin space implies that the four probabilities $p_{3/2,M;2,1,1,1}$ (corresponding to different values of $M$) coincide.
Analogous terms are obtained for ${\bf n} = (1,2,1,1)$, ${\bf n} = (1,1,2,1)$, and ${\bf n} = (1,1,1,2)$, simply by exchanging the index $i$ with either $j$, $k$, or $l$.

\subsubsection{Six electrons}
\paragraph{\color{black} Subspace $S=0$.}
The six-electron component with three singlets is defined in subspaces such as $\zeta=(0,0;2,2,1,1)$, spanned by the state $|\mathcal{S}_{ii},\mathcal{S}_{jj},\mathcal{S}_{kl}\rangle$. 
The generic density operator within such subspace can be written as: 
\begin{align}
|\mathcal{S}_{ij} \rangle\langle \mathcal{S}_{ij}| \otimes \hat{\mathcal{P}}_{k|2}\,\hat{\mathcal{P}}_{l|2}\,, 
\end{align}
corresponding to a projector on two localized and a delocalized singlet state. 

\paragraph{\color{black} Subspace $S=1$.}
The six-electron component with two singlets is defined in subspaces such as $\zeta=(1,M;2,2,1,1)$. 
The generic density operator within this subspace can be written as a combination of terms: 
\begin{align}
\hat{\rho}_{1,M;2,2,1,1}^{(4)} =  \hat{\mathcal{P}}_{i|2}\,\hat{\mathcal{P}}_{j|2}\,\hat{\mathcal{P}}_{kl|1,M} \,.
\end{align}

Analogous terms are obtained for ${\bf n} = (2,1,2,1)$, ${\bf n} = (2,1,1,2)$, ${\bf n} = (1,2,2,1)$, ${\bf n} = (1,2,1,2)$, and ${\bf n} = (1,1,2,2)$, simply by exchanging the indices $i$ and/or $j$ with $k$ and/or $l$.

\subsubsection{Seven electrons}
\paragraph{\color{black} Subspace $S=1/2$.}
The seven-electron terms are defined in subspaces such as $\zeta=[1/2,M,(2,2,2,1)]$ spanned by the states $|\mathcal{S}_{ii},\mathcal{S}_{jj},\mathcal{S}_{kk},\alpha_l\rangle$, characterized by three localized singlets and one unpaired electron. 
The generic density operator within the resulting two-dimensional subspace $S=1/2$ can be written as a combination, with equal weights, of the two terms ($M=\pm 1/2$): 
\begin{align}
\hat{\rho}_{1/2,M;2,2,2,1}^{(4)} =  \hat{\mathcal{P}}_{i|2}\,\hat{\mathcal{P}}_{j|2}\,\hat{\mathcal{P}}_{k|2}\,\hat{\mathcal{P}}_{l|1/2,M} \,.
\end{align}
Analogous terms are obtained for ${\bf n} = (2,2,1,2)$, ${\bf n} = (2,1,2,2)$, and ${\bf n} = (1,2,2,2)$, simply by exchanging the index $l$ with $i$, $j$, or $k$..

\subsubsection{Eight electrons}
\paragraph{\color{black} Subspace $S=0$.}
The eight-electron term coincides with the projector:
\begin{align}
\hat{\rho}_{0,0;2,2,2,2}^{(4)} = \hat{\mathcal{P}}_{i|2}\,\hat{\mathcal{P}}_{j|2}\,\hat{\mathcal{P}}_{k|2}\,\hat{\mathcal{P}}_{l|2}\,,
\end{align}
related to the product of four localized singlets.

}

\subsection{Beyond four-mode subsystems\label{subsecE}}

\begin{table}[]
\centering
\begin{tabular}{|c|cc|c|}
\hline
\multicolumn{4}{|c|}{${\bf n}=(1,1,1,1,1)\ \ \ [z(5,5)=1]$ }\\
\hline
Subspace \# & $S$ & $d_S$ & State $M\!=\!S$ (example) \\
\hline
1 & $1/2$ & 5 & $|\mathcal{S}_{ij},\mathcal{S}_{kl},\Uparrow_m\rangle$ \\
2 & $3/2$ & 4 & $|\mathcal{S}_{ij},\Uparrow_k,\Uparrow_l,\Uparrow_m\rangle$ \\
3 & $5/2$ & 1 & $|\Uparrow_i,\Uparrow_j,\Uparrow_k,\Uparrow_l,\Uparrow_m\rangle$ \\
\hline
\hline
\multicolumn{4}{|c|}{${\bf n}=(2,1,1,1,1)\ \ \ [z(5,6)=5]$ }\\
\hline
Subspace \# & $S$ & $d_S$ & State $M\!=\!S$ (example) \\
\hline
4-8 & $0$ & 2 & $|\mathcal{S}_{ii},\mathcal{S}_{jk},\mathcal{S}_{lm}\rangle$ \\
9-13 & $1$ & 3 & $|\mathcal{S}_{ii},\mathcal{S}_{jk},\Uparrow_l,\Uparrow_m\rangle$ \\
14-18 & $2$ & 1 & $|\mathcal{S}_{ii},\Uparrow_j,\Uparrow_k,\Uparrow_l,\Uparrow_m\rangle$ \\
\hline
\hline
\multicolumn{4}{|c|}{${\bf n}=(2,2,1,1,1)\ \ \ [z(5,7)=10]$ }\\
\hline
Subspace \# & $S$ & $d_S$ & State $M\!=\!S$ (example) \\
\hline
19-28 & $1/2$ & 2 & $|\mathcal{S}_{ii},\mathcal{S}_{jj},\mathcal{S}_{kl},\Uparrow_m\rangle$ \\
29-38 & $3/2$ & 1 & $|\mathcal{S}_{ii},\mathcal{S}_{jj},\Uparrow_k,\Uparrow_l,\Uparrow_m\rangle$ \\
\hline
\hline
\multicolumn{4}{|c|}{${\bf n}=(2,2,2,1,1)\ \ \ [z(5,8)=10]$ }\\
\hline
Subspace \# & $S$ & $d_S$ & State $M\!=\!S$ (example) \\
\hline
39-48 & $0$ & 1 & $|\mathcal{S}_{ii},\mathcal{S}_{jj},\mathcal{S}_{kk},\mathcal{S}_{lm}\rangle$ \\
49-58 & $1$ & 1 & $|\mathcal{S}_{ii},\mathcal{S}_{jj},\mathcal{S}_{kk},\Uparrow_l,\Uparrow_m\rangle$ \\
\hline
\hline
\multicolumn{4}{|c|}{${\bf n}=(2,2,2,2,1)\ \ \ [z(5,9)=5]$ }\\
\hline
Subspace \# & $S$ & $d_S$ & State $M\!=\!S$ (example) \\
\hline
59-63 & $1/2$ & 1 & $|\mathcal{S}_{ii},\mathcal{S}_{jj},\mathcal{S}_{kk},\mathcal{S}_{ll},\Uparrow_m\rangle$ \\
\hline
\hline
\multicolumn{4}{|c|}{${\bf n}=(2,2,2,2,2)\ \ \ [z(5,10)=1]$ }\\
\hline
Subspace \# & $S$ & $d_S$ & State $M\!=\!S$ (example) \\
\hline
64 & $0$ & 1 & $|\mathcal{S}_{ii},\mathcal{S}_{jj},\mathcal{S}_{kk},\mathcal{S}_{ll},\mathcal{S}_{mm}\rangle$ \\
\hline
\end{tabular}
\caption{\label{TableIII} Subspaces corresponding to five modes (orbitals $\phi_i$, $\phi_j$, $\phi_k$, $\phi_l$, and $\phi_m$). Each subspace is defined by the total spin $S$, its projection $M=S$, and by the site occupation numbers ${\bf n}$. The numbers $d_S$ and $z(n,N_e)$ are the multiplicity related to scalar quantum numbers and the number of possible mode occupations ${\bf n}$. The overall numbers of subspaces that block diagonalize the RDO and the 5BRDM, including the multiplicity {\color{black}$(2S+1)$} related to the total spin projection, are $A=168$ and $B=12$.}
\end{table}

In general, the identification of the subspaces that block diagonalize the RDOs, based on the mode occupation numbers, the total spin $S$ and its projection $M$, proceeds as follows. 

The first set of blocks we consider is related to $N_e=n$ particles, where $n$ is the number of modes: in fact, for $N_e<n$ there must be at least one unoccupied mode, and the component of the RDO does not contribute to the $n$BRDM. For even (odd) values of $n$ the total spin can range from $0$ ($1/2$) to $n/2$. For each value of $S$, one has $2S+1$ subspaces, corresponding to different values of $M$, with $-S \le M \le S$. Each of these has a multiplicity $d_S$, related to scalar quantum numbers, such as the partial spin sums \cite{Tsukerblat}. For each subspace with $M=S$, the density operator can be expressed in terms of an overcomplete set of states, defined by the product of $2S$ unpaired spins in the state $\Uparrow$ and of $n-2S$ spins forming $n/2-S$ delocalized singlets. Due to rotational invariance, the density operator within a subspace with $M<S$ can be obtained from that with $M=S$ simply by replacing, in each state of the overcomplete basis, the $|\!\Uparrow\dots\Uparrow\rangle$ state of the unpaired electrons with {\color{black} the totally symmetric one} corresponding to the same $S$ and to the relevant $M$ \cite{Karbach93a}. For example, the four-mode ($n=4$) state $|\mathcal{S}_{12},\Uparrow_3,\Uparrow_4\rangle$, belonging $M=S=1$, corresponds to $\frac{1}{\sqrt{2}} (|\mathcal{S}_{12},\Uparrow_3,\Downarrow_4\rangle + |\mathcal{S}_{12},\Downarrow_3,\Uparrow_4\rangle)$ in the subspace $M=0$. 

Other sets of blocks are related to $N_e>n$ particles. The relevant terms for the derivation of the NBDRMs are those corresponding to singly and doubly occupied modes only. This subspace is further partitioned based on the set of modes that are doubly occupied. In fact, as already mentioned, the RDOs cannot have off-diagonal terms between states with different mode occupations (${\bf n}$). Each subspace with given  mode occupations can be further divided in subspaces, based on the values of $S$ and $M$. Here, one can apply the same procedure discussed above for $N_e=n$, by simply {\color{black} replacing $n$ with $2n-N_e$, as} the number of {\color{black} relevant ({\it i.e.} singly-occupied)} modes. 

Overall, the number of subspaces that block diagonalize the RDO of a subsystem formed by $n$ modes is given by
\begin{align}
    A = \sum_{N_e=n}^{2n} \left( \begin{array}{c}
        n \\ N_e-n
    \end{array} \right) \sum_{S=S_{\rm min}}^{S_{\rm max}} (2S+1)\,.
\end{align}
Here, $S_{\rm min}=0$ or $1/2$ depending on whether {\color{black} the number of singly-occupied orbitals $q=2n-N_e$} is even or odd {\color{black} $S_{\rm max}=n-N_e/2$}; the binomial coefficient (corresponding to the $z(n,N_e)$ given in the Tables \ref{TableI}-\ref{TableIV}) counts the different arrangements of the $N_e-n$ localized singlets within the $n$ modes; the last factor accounts for the multiplicty related to $M$ within each $S$ multiplet. 

Finally, $d_S$ is the multiplicity of each $S$ multiplet related to the scalar quantities, such as the partial spin sums. For even numbers $q=2n-N_e$ of singly occupied modes, this is given by \cite{Karbach93a}:
\begin{align}
    d_S (q) = 
\frac{q!\, (2S+1)}{(q/2-S)!\,(S+q/2+1)!}
\end{align}
and the number of states that form the overcomplete basis reads
\begin{align}
    w = \frac{q!}{2^{q/2} (q/2)!}\,.
\end{align}
For odd values $q$ of singly occupied modes, one has that:
\begin{align}
    d_S(q) = d_{S-1/2} (q-1) + d_{S+1/2} (q-1) 
\end{align}
and the number of states that form the overcomplete basis reads
\begin{align}
    w = \frac{q!}{2^{(q-1)/2} [(q-1)/2]!}\,.
\end{align}

The above procedure for identifying the subspaces that block diagonalize the RDOs is applied to the cases of five and six modes (Tables \ref{TableIII} and \ref{TableIV}, respectively). In these tables, like in those reported in the previous Subsections for three and four modes (Tables \ref{TableI} and \ref{TableII}), we list the subspaces corresponding to different particle numbers. In order to exemplify, we report a specific set of double occupations, specified by the vector ${\bf n}$, among the $z$ possible ones. Each subspace with given ${\bf n}$, $S$ and $M$ has a dimension $d_S$ and is spanned by an overcomplete basis formed by $w$ vectors, each one given by the product of $N_e-n$ localized and $n-N_e/2-S$ delocalized spin singlets, and of $2S$ parallel spins. In the tables, we report for simplicity one vector belonging to the subspace $M=S$. The vectors {\color{black} belonging to the subspace with equal $S$ and ${\bf n}$, but different $M<S$, are obtained from the $M=S$ state by replacing the polarized state of the $2S$ spins with the fully symmetric state corresponding to the relevant value of $M$.}

\begin{table}[]
\centering
\begin{tabular}{|c|cc|c|}
\hline
\multicolumn{4}{|c|}{${\bf n}=(1,1,1,1,1,1)\ \ \ [z(6,6)=1]$ }\\
\hline
Subspace \# & $S$ & $d_S$ & State $M\!=\!S$ (example) \\
\hline
1 & $0$ & 5 & $|\mathcal{S}_{ij},\mathcal{S}_{kl},\mathcal{S}_{mn}\rangle$ \\
2 & $1$ & 9 & $|\mathcal{S}_{ij},\mathcal{S}_{kl},\Uparrow_m,\Uparrow_n\rangle$ \\
3 & $2$ & 5 & $|\mathcal{S}_{ij},\Uparrow_k,\Uparrow_l,\Uparrow_m,\Uparrow_n\rangle$ \\
4 & $3$ & 1 & $|\Uparrow_i,\Uparrow_j,\Uparrow_k,\Uparrow_l,\Uparrow_m,\Uparrow_n\rangle$ \\
\hline
\hline
\multicolumn{4}{|c|}{${\bf n}=(2,1,1,1,1,1)\ \ \ [z(6,7)=6]$ }\\
\hline
Subspace \# & $S$ & $d_S$ & State $M\!=\!S$ (example) \\
\hline
5-10 & $1/2$ & 5 & $|\mathcal{S}_{ii},\mathcal{S}_{jk},\mathcal{S}_{lm},\Uparrow_n\rangle$ \\
11-16 & $3/2$ & 4 & $|\mathcal{S}_{ii},\mathcal{S}_{jk},\Uparrow_l,\Uparrow_m,\Uparrow_n\rangle$ \\
17-22 & $5/2$ & 1 & $|\mathcal{S}_{ii},\Uparrow_j,\Uparrow_k,\Uparrow_l,\Uparrow_m,\Uparrow_n\rangle$ \\
\hline
\hline
\multicolumn{4}{|c|}{${\bf n}=(2,2,1,1,1,1)\ \ \ [z(6,8)=15]$ }\\
\hline
Subspace \# & $S$ & $d_S$ & State $M\!=\!S$ (example) \\
\hline
23-37 & $0$ & 2 & $|\mathcal{S}_{ii},\mathcal{S}_{jj},\mathcal{S}_{kl},\mathcal{S}_{mn}\rangle$ \\
38-52 & $1$ & 3 & $|\mathcal{S}_{ii},\mathcal{S}_{jj},\mathcal{S}_{kl},\Uparrow_m,\Uparrow_n\rangle$ \\
53-67 & $2$ & 1 & $|\mathcal{S}_{ii},\mathcal{S}_{jj},\Uparrow_k,\Uparrow_l,\Uparrow_m,\Uparrow_n\rangle$ \\
\hline
\hline
\multicolumn{4}{|c|}{${\bf n}=(2,2,2,1,1,1)\ \ \ [z(6,9)=20]$ }\\
\hline
Subspace \# & $S$ & $d_S$ & State $M\!=\!S$ (example) \\
\hline
68-87 & $1/2$ & 2 & $|\mathcal{S}_{ii},\mathcal{S}_{jj},\mathcal{S}_{kk},\mathcal{S}_{lm},\Uparrow_n\rangle$  \\
88-107 & $3/2$ & 1 & $|\mathcal{S}_{ii},\mathcal{S}_{jj},\mathcal{S}_{kk},\Uparrow_l,\Uparrow_m,\Uparrow_n\rangle$ \\
\hline
\hline
\multicolumn{4}{|c|}{${\bf n}=(2,2,2,2,1,1)\ \ \ [z(6,10)=15]$ }\\
\hline
Subspace \# & $S$ & $d_S$ & State $M\!=\!S$ (example)  \\
\hline
108-122 & $0$ & 1 & $|\mathcal{S}_{ii},\mathcal{S}_{jj},\mathcal{S}_{kk},\mathcal{S}_{ll},\mathcal{S}_{mn}\rangle$  \\
123-137 & $1$ & 1 & $|\mathcal{S}_{ii},\mathcal{S}_{jj},\mathcal{S}_{kk},\mathcal{S}_{ll},\Uparrow_m,\Uparrow_n\rangle$  \\
\hline
\hline
\multicolumn{4}{|c|}{${\bf n}=(2,2,2,2,2,1)\ \ \ [z(6,11)=6]$ }\\
\hline
Subspace \# & $S$ & $d_S$ & State $M\!=\!S$ (example)  \\
\hline
138-143 & $1/2$ & 1 & $|\mathcal{S}_{ii},\mathcal{S}_{jj},\mathcal{S}_{kk},\mathcal{S}_{ll},\mathcal{S}_{mm},\Uparrow_n\rangle$  \\
\hline
\hline
\multicolumn{4}{|c|}{${\bf n}=(2,2,2,2,2,2)\ \ \ [z(6,12)=1]$ }\\
\hline
Subspace \# & $S$ & $d_S$ & State $M\!=\!S$ (example)  \\
\hline
144 & $0$ & 1 & $|\mathcal{S}_{ii},\mathcal{S}_{jj},\mathcal{S}_{kk},\mathcal{S}_{ll},\mathcal{S}_{mm},\mathcal{S}_{nn}\rangle$ \\
\hline
\end{tabular}
\caption{\label{TableIV} Subspaces corresponding to six modes ($\phi_i$, $\phi_j$, $\phi_k$, $\phi_l$, $\phi_m$, and $\phi_n$). Each subspace is defined by the total spin $S$, its projection $M=S$, and by the site occupation numbers ${\bf n}$. The numbers $d_S$ and $z(n,N_e)$ are the multiplicity related to scalar quantum numbers and the number of possible mode occupations ${\bf n}$, respectively. The overall numbers of subspaces that block diagonalize the RDO and the 6BRDM, including the multiplicity {\color{black} $(2S+1)$} related to the total spin projection, are $A=416$ and $B=16$.}
\end{table}

\section{$n$-body reduced density matrices\label{sec4}}

\begin{figure}
\centering
\includegraphics[width=0.4\textwidth]{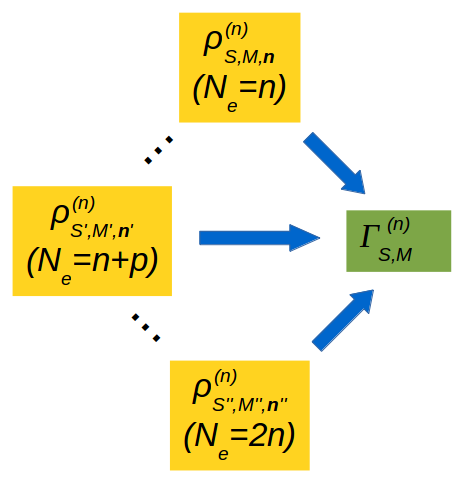}
\caption{\color{black}
Different terms of the $n$-mode RDO contribute to each term $\hat{\Gamma}^{n}_{S;M}$ of the $n$BRDM.
The first contribution (top) comes from the term $\hat{\rho}^{(n)}_\zeta$ with identical values of $S$ and $M$, and no double occupations [${\bf n}=(1,\dots,1)$]. Further contributions (left) come from terms $\hat{\rho}^{(n)}_{S',M';{\bf n}}$, characterized by $p$ doubly-occupied modes and values of the spin quantum numbers such that $|S-S'|,|M-M'| \le p/2$. The term of the RDO (bottom) where all orbitals are doubly occupied ($p=n$, $S''=M''=0$) contributes to all the $\hat{\Gamma}^{n}_{S;M}$ (with $S,M \le n/2$).
}
\label{fig3}
\end{figure}

Given the expressions of the RDOs, one can derive those of the $n$BRDMs. In particular, we are interested in the $n$BRDMs that are diagonal with respect to the mode label, while including all the coherences between the spin degrees of freedom. The $n$BRDMs can be blocked diagonalized in terms of subspaces defined by the values of $S$ and $M$. 

Paralleling the structure of the previous Section, we start discussing the general properties of the $n$BRDMs (Subsec. \ref{sbsgp}). Then, we report - for illustrative purposes - the cases of $n=1$ (Subsec. \ref{sbs1b}) and $n=2$ (Subsec. \ref{sbs2b}). Finally, we discuss in detail the cases of $n=3$ (Subsec. \ref{sbs3b}) and $n=4$ (Subsec. \ref{sbs4b}).

{

\subsection{General properties\label{sbsgp}}

The $n$BRDM has the same structure as the RDO. In particular, it formally corresponds to the block with $N_e=n$ particles, where all modes are singly occupied. It can thus be divided in blocks (subspaces), each one characterized by well-defined values of $S$ and $M$: the former one varies from $0$ or $1/2$ to $n/2$, the latter one from $-S$ to $S$. The numbers of these subspaces are the same ones discussed for the RDO with $N_e=n$ particles (see Tables \ref{TableI}-\ref{TableIV}), and are given by:
\begin{align}
    B = \sum_{S=S_{\rm min}}^{S_{\rm max}} (2S+1)\,,
\end{align}
where $S_{\min}=0$ or $1/2$, depending on whether $n$ is even or odd{\color{black}, and $S_{\rm max}=n/2$}. 

The $n$BRDM results from different terms of the RDO (see Fig.~\ref{fig3}). One contribution comes from the block with $N_e=n$, which can be directly mapped onto the $n$BRDM: the block of the RDO corresponding to given values, of the total spin and of its projection contributes to the block of the $n$BRDM with equal values of $S$ and $M$. 

If $N_e>n$, the subspace contains $N_e-n$ localized singlets. These give rise to a mixture of spin up and spin down states in the $n$BRDM. In fact, the connection between each term in the RDO and the corresponding one in the $n$BRDM can be easily specified in terms of the projectors that enter the expressions of the terms $\hat{\rho}^{(n)}_\zeta$:
$\hat{\mathcal{P}}_{i|\Uparrow} \!\rightarrow\! \hat{P}_{i|\uparrow}$,
$\hat{\mathcal{P}}_{i|\Downarrow} \!\rightarrow\! \hat{P}_{i|\downarrow}$,
$|\mathcal{S}_{ij} \rangle\langle \mathcal{S}_{ij}| \!\rightarrow\! |S_{ij} \rangle\langle S_{ij}|$,
and 
$\hat{\mathcal{P}}_{i|2} \!\rightarrow\! \hat{P}_{i|\uparrow} \!+\! \hat{P}_{i|\downarrow}$. {
Here, we remind that $\Uparrow$, $\Downarrow$, and $\mathcal{S}$ denote fermionic states (expressed in the second quantization formalism), while $\uparrow$, $\downarrow$, and $S$ denote spin states. Analogously, $\hat{\mathcal{P}}$ and $\hat{P}$ are projector operators that act in the fermionic and in the spin spaces, respectively.}
From the above it follows that each term $\hat{\rho}^{(n)}_{S,M;{\bf n}}$ with $p$ doubly occupied modes ($N_e=n+p$) can contribute to the components of the $n$BRDM with total spin $S'$ and total spin projection $M'$ such that $|S-S'|\le p/2$ and $|M-M'|\le p/2$.
From this and from the equations reported in the previous Section one can derive the expressions of the $n$BRDMs, whose contributions can then be classified based on the quantum numbers $S$ and $M$.

\subsection{One-body reduced density matrix\label{sbs1b}}

The one-body reduced density matrix referred to mode $i$ is defined as 
\begin{align}
    \Gamma_{i;\sigma_i,\sigma_i'} = \langle \Psi | \hat{d}_{i,\sigma_i'}^\dagger\, \hat{d}_{i,\sigma_i} |\Psi\rangle = {\rm Tr} \left(\hat{\rho}_i\, \hat{d}_{i,\sigma_i'}^\dagger\, \hat{d}_{i,\sigma_i}\right),
\end{align}
where $\hat{\rho}_i$ is the single-mode reduced density operator defined in Subsec. \ref{subsecA}. 

The expression of the 1BRDM can be derived from that of the RDO $\hat{\rho}_i$ and is divided into contributions corresponding to specific values of $M$, for $S=1/2$:
\begin{align}
    \hat{\Gamma}^{(1)} = \sum_{M=-1/2}^{+1/2} \hat{\Gamma}^{(1)}_{1/2,M}
\end{align}
where the two contributions read
\begin{align}
    \hat{\Gamma}^{(1)}_{1/2,M} = \left[ p^{(1)}_{1/2,M;1} + p^{(1)}_{0,0;2}\right] \hat{\rho}^{(1)}_{1/2,M;1}\,.
\end{align}
The first and second contributions above originate from the single and double occupations of the orbital $\phi_i$, respectively. Due to rotational invariance, $0\le p^{(1)}_{1/2,1/2;2}\le 1/2$ (because of the constraint $p^{(1)}_{1/2,1/2;2} = p^{(1)}_{1/2,-1/2;2}$), whereas $0\le p^{(1)}_{0,0;2}\le 1$. Analogously, in all the following expressions of the present Section, the probabilities are bounded by the inequalities: 
\begin{align}
0 \le p^{(n)}_{S,M;{\bf n}} \le 1/(2S+1) \,.
\end{align}

\subsection{Two-body reduced density matrix\label{sbs2b}}

The two-body reduced density matrix referred to the modes $i$ and $j$ is given by the expression 
\begin{align}
    \Gamma_{ij;{\bf\sigma},{\bf\sigma}'} & = \langle \Psi | \hat{d}_{j,\sigma_j'}^\dagger\, \hat{d}_{i,\sigma_i'}^\dagger\, \hat{d}_{i,\sigma_i}\, \hat{d}_{j,\sigma_j} |\Psi\rangle \nonumber\\ & = {\rm Tr} \left(\hat{\rho}_{ij}\, \hat{d}_{j,\sigma_j'}^\dagger\, \hat{d}_{i,\sigma_i'}^\dagger\, \hat{d}_{i,\sigma_i}\, \hat{d}_{j,\sigma_j}\right),
\end{align}
where ${\bf\sigma}\equiv (\sigma_i,\sigma_j)$, ${\bf\sigma}'\equiv (\sigma_i',\sigma_j')$, and $\hat{\rho}_{ij}$ is the two-mode reduced density operator defined in Subsec.~\ref{subsecB}. 

The expression of the 2BRDM can be derived from that of the RDO and is divided into contributions corresponding to specific values of $S$ and $M$:
\begin{align}
    \hat{\Gamma}^{(2)} = \sum_{S=0}^1 \sum_{M=-S}^{+S} \hat{\Gamma}^{(2)}_{S,M}\,.
\end{align}
The different contributions are given by 
\begin{align}
    \hat{\Gamma}^{(2)}_{S,M} & = p^{(2)}_{S,M;1,1} \, \hat{\rho}^{(2)}_{S,M;1,1}
\nonumber\\ & \quad + \left[ \sum_{\bf n}' p^{(2)}_{1/2,1/2;{\bf n}} + p^{(2)}_{0,0;2,2} \right]\hat{P}_{S,M}\,,\label{eq:x1}
\end{align}
where the sum over ${\bf n}$ includes the cases $(2,1)$ and $(1,2)$, and $\hat{P}_{S,M}$ is the projector on the subspace corresponding to one spin in each of the two modes $i$ and $j$, with total spin $S$ and projection $M$.

From the above equation it follows that an excess singlet probability can only come from the two-particle contribution in the reduced density operator $\hat{\rho}_{ij}$ [first line in Eq.~\eqref{eq:x1}], whereas the three- and four-particle contributions (second line) contribute equally to the singlet and to the triplet states, the expression in square parentheses being independent on $S$ and $M$.

\subsection{Three-body reduced density matrix\label{sbs3b}}

The three-body reduced density matrix referred to modes $i$, $j$ and $k$, is given by the expression 
\begin{align}
    &\Gamma_{ijk;{\bf\sigma},{\bf\sigma'}} = \langle \Psi | \hat{d}_{k,\sigma_k'}^\dagger\, \hat{d}_{j,\sigma_j'}^\dagger\, \hat{d}_{i,\sigma_i'}^\dagger\, \hat{d}_{i,\sigma_i}\, \hat{d}_{j,\sigma_j}\, \hat{d}_{k,\sigma_k} |\Psi\rangle \nonumber\\ & \quad = {\rm Tr} \left(\hat{\rho}_{ijk}\, \hat{d}_{k,\sigma_k'}^\dagger\, \hat{d}_{j,\sigma_j'}^\dagger\, \hat{d}_{i,\sigma_i'}^\dagger\, \hat{d}_{i,\sigma_i}\, \hat{d}_{j,\sigma_j}\, \hat{d}_{k,\sigma_k}\right),
\end{align}
where ${\bf\sigma}\equiv (\sigma_i,\sigma_j,\sigma_k)$,  ${\bf\sigma}'\equiv (\sigma_i',\sigma_j',\sigma_k')$, and $\hat{\rho}_{ijk}$ is the three-mode reduced density operator defined in Subsec.~\ref{subsecC}. 

The expression of the 3BRDM can be derived from that of the RDO and is divided into contributions corresponding to specific values of $S$ and $M$:
\begin{align}
    \hat{\Gamma}^{(3)} = \sum_{S=1/2}^{3/2} \sum_{M=-S}^{+S} \hat{\Gamma}^{(3)}_{S,M}\,.
\end{align}
The different contributions within the $S=1/2$ subspace are given by
\begin{align}
    \hat{\Gamma}^{(3)}_{1/2,M} & = 
    p^{(3)}_{1/2,M;1,1,1} \hat{\rho}^{(3)}_{1/2,M;1,1,1} \nonumber\\
    & + \sum_{\bf n}' p^{(3)}_{0,0;{\bf n}} \hat{P}_{1/2,M,S_{ab}=0}  
    + \sum_{\bf n}' p^{(3)}_{1,1;{\bf n}} \hat{P}_{1/2,M,S_{ab}=1} \nonumber\\
    & + \left[ \sum_{\bf n}'' p^{(3)}_{1/2,1/2;{\bf n}}   
    + p^{(3)}_{0,0;2,2,2} \right] \hat{P}_{1/2,M}\,,\label{eq:x2}
\end{align}
while those within the $S=3/2$ subspace read
\begin{align}
    \hat{\Gamma}^{(3)}_{3/2,M} & = 
    p^{(3)}_{3/2,M;1,1,1} \hat{\rho}^{(3)}_{3/2,M;1,1,1} 
    + \sum_{\bf n}' p^{(3)}_{1,1;{\bf n}} \hat{P}_{3/2,M}  \nonumber\\
     & \quad + \left[  
    \sum_{\bf n}'' p^{(3)}_{1/2,1/2;{\bf n}} + p^{(3)}_{0,0;2,2,2}\right] \hat{P}_{3/2,M}\,.\label{eq:x3}
\end{align}
Here, the sum $\sum'$ includes the cases $(2,1,1)$, $(1,2,1)$, and $(1,1,2)$, with $a$ and $b$ the two singly-occupied modes;
the sum $\sum''$ includes the cases $(2,2,1)$, $(2,1,2)$, and $(1,2,2)$.
In Eq.~\eqref{eq:x2}, $\phi_a$ and $\phi_b$ are the singly-occupied orbitals, according to the current vector ${\bf n}$.
From the above expressions it follows that the difference between the doublet ($S=1/2$) and the quadruplet ($S=3/2$) terms is induced by the three-particle contribution to the reduced density operator $\hat{\rho}_{ijk}$ [first term in Eqs.~(\ref{eq:x2},\ref{eq:x3})], and from the four-particle contribution corresponding to a zero partial-spin sum [second term in Eq.~\eqref{eq:x2}]. The five- and six-particle, instead, terms give equal contributions to $\hat{\Gamma}^{(3)}_{1/2,M}$ and $\hat{\Gamma}^{(3)}_{3/2,M}$.

Overall, the 3BRDM can be written in a block-diagonal form with respect to the six subspaces, which are defined by the values of $S$ and $M$ [see the first block, corresponding to ${\bf n}=(1,1,1)$, of Table \ref{TableI}, and apply the replacements: $\Uparrow\rightarrow\uparrow$, $\Downarrow\rightarrow\downarrow$, $\mathcal{S}\rightarrow S$].

If the $n$BRDM only contains real elements, then it can be written as a combination of projectors on products of singlet states ($ | S_{ij} \rangle\langle S_{ij} | $) and on single spin subspaces ($\hat{P}_{i|\uparrow} + \hat{P}_{i|\downarrow}$). In particular, as reported in Ref.~\cite{Lunkes05a} for the case where the particle extraction takes place from a Fermi gas at $T=0\,$K and at defined positions, the $n$BRDM takes the form:
\begin{align}
    \hat{\Gamma}^{(3)} = \sum_{ab=ij,jk,ki} \frac{p_{ab}}{2} | S_{ab} \rangle\langle S_{ab} | \otimes P_c + \frac{p_0}{8} P_i P_j P_k \,,
\end{align}
where $P_c \equiv |\!\uparrow_c\rangle\langle\uparrow_c\!| + |\!\downarrow_c\rangle\langle\downarrow_c\!|$, $c=i,j,k$ and $c\neq a,b$. It should be noted that the coefficients $p_{ab}$ and $p_0$ sum to 1, but can also take negative values, so that the above expression can also correspond to a genuine multipartite entangled state \cite{Vertesi07a}.

\subsection{Four-body reduced density matrix\label{sbs4b}}

The four-body reduced density matrix referred to the modes $i$, $j$, $k$ and $l$, is given by the expression 
\begin{align}
    &\Gamma_{ijkl;{\bf\sigma},{\bf\sigma'}} \!=\! \langle \Psi | \hat{d}_{l,\sigma_l'}^\dagger \hat{d}_{k,\sigma_k'}^\dagger \hat{d}_{j,\sigma_j'}^\dagger \hat{d}_{i,\sigma_i'}^\dagger \hat{d}_{i,\sigma_i} \hat{d}_{j,\sigma_j} \hat{d}_{k,\sigma_k} \hat{d}_{l,\sigma_l} |\Psi\rangle \nonumber\\ &= {\rm Tr} \left(\rho_{ijkl} \hat{d}_{l,\sigma_l'}^\dagger \hat{d}_{k,\sigma_k'}^\dagger \hat{d}_{j,\sigma_j'}^\dagger \hat{d}_{i,\sigma_i'}^\dagger \hat{d}_{i,\sigma_i} \hat{d}_{j,\sigma_j} \hat{d}_{k,\sigma_k} \hat{d}_{l,\sigma_l} \right),
\end{align}
where ${\bf\sigma}\equiv (\sigma_i,\sigma_j,\sigma_k,\sigma_l)$,  ${\bf\sigma}'\equiv (\sigma_i',\sigma_j',\sigma_k',\sigma_l')$, and $\rho_{ijkl}$ is the four-mode reduced density operator defined in Subsec.~\ref{subsecD}. 

The expression of the 4BRDM can be derived from that of the RDO and are divided into contributions corresponding to specific values of $S$ and $M$:
\begin{align}
    \hat{\Gamma}^{(4)} = \sum_{S=0}^{2} \sum_{M=-S}^{+S} \hat{\Gamma}^{(4)}_{S,M}\,.
\end{align}
The different contributions for $S=0$ are given by
\begin{widetext}
\begin{align}
\hat{\Gamma}^{(4)}_{0,0} & = 
p^{(4)}_{0,0;1,1,1,1} \hat{\rho}^{(4)}_{0,0;1,1,1,1} + 
\left[ \sum_{\bf n}''' p^{(4)}_{1/2,1/2;{\bf n}} + p^{(4)}_{0,0;2,2,2,2} \right] \hat{P}_{0,0} \nonumber\\
& \quad + \sum_{\bf n}' p^{(4)}_{1/2,1/2;{\bf n}} \hat{P}_{0,0,S_{abc}=1/2} + \sum_{\bf n}'' p^{(4)}_{0,0;{\bf n}} \hat{P}_{0,0,S_{ab}=0} 
+ \sum_{\bf n}'' p^{(4)}_{1,1;{\bf n}} \hat{P}_{0,0,S_{ab}=1}\,,\label{eq:x4}
\end{align}
those for $S=1$ by
\begin{align}
\hat{\Gamma}^{(4)}_{1,M} & = 
p^{(4)}_{1,M;1,1,1,1} \hat{\rho}^{(4)}_{1,M;1,1,1,1}  + 
\left[ \sum_{\bf n}''' p^{(4)}_{1/2,1/2;{\bf n}} + p^{(4)}_{0,0;2,2,2,2} \right] \hat{P}_{1,M} 
+ \sum_{\bf n}' p^{(4)}_{1/2,1/2;{\bf n}} \hat{P}_{1,M,S_{abc}=1/2} 
\nonumber\\
& \quad + \sum_{\bf n}' p^{(4)}_{3/2,3/2;{\bf n}} \hat{P}_{1,M,S_{abc}=3/2} 
+ \sum_{\bf n}'' \left[ p^{(4)}_{0,0;{\bf n}} \hat{P}_{1,M,S_{ab}=0} 
+ p^{(4)}_{1,1;{\bf n}} \hat{P}_{1,M,S_{ab}=1}\right] \,,\label{eq:x5}
\end{align}
and those for $S=2$ by
\begin{align}
\hat{\Gamma}^{(4)}_{2,M} = 
p^{(4)}_{2,M;1,1,1,1} \hat{\rho}^{(4)}_{2,M;1,1,1,1} + 
\left[\sum_{\bf n}' p^{(4)}_{3/2,3/2;{\bf n}}+ \sum_{\bf n}'' p^{(4)}_{1,1;{\bf n}}  
+ \sum_{\bf n}''' p^{(4)}_{1/2,1/2;{\bf n}} + p^{(4)}_{0,0;2,2,2,2}\right]\hat{P}_{2,M}\,.\label{eq:x6}
\end{align}
\end{widetext}
Here, the sums $\sum'$, $\sum''$, and $\sum'''$ include all the ${\bf n}$ with only one, two, and three doubly occupied modes, respectively. 
In Eqs.~(\ref{eq:x4}-\ref{eq:x5}), $\phi_a$, $\phi_b$, and $\phi_c$ are the singly-occupied orbitals, according to the current vector ${\bf n}$.
The difference between the singlet ($S=0$), triplet ($S=1$), and quintuplet ($S=2$) contributions results from the four-, five-, and six-particle terms in $\hat{\rho}_{ijkl}$, while the seven- and eight-particle terms contribute equally to $\hat{\Gamma}_{0,0}$, $\hat{\Gamma}_{1,M}$, and $\hat{\Gamma}_{3,M}$.

Overall, the 4BRDM can be written in a block-diagonal form with respect to the nine subspaces, which are defined by the values of $S$ and $M$ [see the first block, corresponding to ${\bf n}=(1,1,1,1)$, of Table \ref{TableII}, and apply the replacements: $\Uparrow\rightarrow\uparrow$, $\Downarrow\rightarrow\downarrow$, $\mathcal{S}\rightarrow S$].

If the $n$BRDM only contains real elements, then it can be written as a combination of projectors on products of singlet states and on single spin states. In contrast with what was reported in Ref.~\cite{Lunkes05a} for the case where the particle extraction takes place from a Fermi gas at $T=0\,$K and at defined positions, the $n$BRDM takes the form:
\begin{align}
    & \hat{\Gamma}^{(4)} = \sum_{ab=ij,ik,il} p_{ab,cd} | S_{ab} , S_{cd} \rangle\langle S_{ab} , S_{cd} | \nonumber\\ 
    & +\! \sum_{ab=ij,ik,il} \frac{p_{ab}}{4} | S_{ab} \rangle\langle S_{ab} | \!\otimes\! P_c P_d \!+\! \frac{p_0}{16} P_i P_j P_k P_l\,,
\end{align}
where $cd=kl,jl,jk$ is the pair complementary to $ab$. Also in this case, the coefficients $p_{ab,cd}$, $p_{ab}$, and $p_0$ sum to 1, but may also take negative values, so that the above expression can also correspond to a genuine multipartite entangled state \cite{Troiani25a}.

{
\section{Implications for spin entanglement\label{sec5}}

Genuine multipartite entanglement can be found in the detected spin states, obtained from a variety of fermionic states \cite{Troiani25a}. The entanglement extracted from cyclic systems resembles that displayed by the ground state of spin rings \cite{Troiani11a,Siloi14a}.  
The amount of such entanglement can however be reduced by the presence of doubly occupied modes $\phi_i$ and by the rotational invariance in subspaces corresponding to $S>0$. These aspects are discussed separately in the following Subsections.

\subsection{The effect of double occupations}

From the equations reported in the previous Sections, it follows that the detected spin states resulting from the charge configurations ${\bf n}$ with double occupations are always separable. More specifically, if the charge configuration contains $p$ doubly occupied detection orbitals $\phi_i$, the resulting state is {\color{black}at least} $(p+1)$-separable. In order to clarify this point, let us consider an $n$-mode RDO
(with mode indices $i,j,\dots=1,2,\dots$)
\begin{align}
\hat{\rho}^{(n)}_\zeta = | S_{11} \rangle\langle S_{11} | \otimes | \psi_{\overline 1}\rangle\langle\psi_{\overline 1}| \,,    
\end{align}
where $| \psi_{\overline 1}\rangle$ is a generic state of the $n-1$ modes other than $1$, all characterized by a single occupation. By applying the definitions provided in Sec. \ref{sec4}, one can show that the corresponding RDO reads
\begin{align}
    \hat{\Gamma}^{(n)} = ( |\!\uparrow_1\rangle\langle\uparrow_1\!\!|+|\!\downarrow_1\rangle\langle\downarrow_1\!\!| ) \otimes | \tilde\psi_{\overline 1}\rangle\langle\tilde\psi_{\overline 1}|\,,
\end{align}
where $| \tilde\psi_{\overline 1}\rangle$ is the purely spin state formally corresponding to the fermionic state $| \psi_{\overline 1}\rangle$ (see Subsec.~\ref{sbsgp}). This state is at least biseparable with respect to the partition $s_1 | \otimes_{i=2}^n s_i$. 

Along the same lines, one can show that an $n$-mode RDO that includes $p$ doubly-occupied orbitals,
\begin{align}\label{xx1}
\hat{\rho}^{(n)}_\zeta = \otimes_{i=1}^p | S_{ii} \rangle\langle S_{ii} | \otimes | \psi_{\overline {1\dots p}}\rangle\langle\psi_{\overline {1\dots p}}| \,,    
\end{align}
gives rise to an $n$BRDM of the form
\begin{align}\label{xx2}
    \hat{\Gamma}^{(n)} = \otimes_{i=1}^p(|\!\uparrow_i\rangle\langle\uparrow_i\!\!|+|\!\downarrow_i\rangle\langle\downarrow_i\!\!|) \otimes | \tilde\psi_{\overline{1 \dots p}} \rangle\langle\tilde\psi_{\overline{1\dots p}}|\,,
\end{align}
{\color{black} where $| \tilde{\psi}_{\overline{1\dots p}}\rangle$ is a generic state of the $n-p$ spins.}
This $n$BRDM corresponds to a spin state that is at least ($p+1$)-separable. 

Three comments are in order here.
First, we note that, in spite of its separability, the $n$BRDM given in Eq.~\eqref{xx2} can display multipartite entanglement within the ($n-p$)-spin state $| \tilde\psi_{\overline{1 \dots p}} \rangle$. Therefore, the presence in the RDO of doubly-occupied orbitals is compatible with the presence of multipartite entanglement in the detected spin state. 
Second, terms like the one reported in Eq.~\eqref{xx1} will generally appear in the expression of the RDO together with other contributions. If all of these include doubly-occupied orbitals, then the resulting $n$BRDM cannot display genuine multipartite entanglement. In fact, the $n$BRDM will consist of a mixture of different terms, each one corresponding to a different charge configuration ${\bf n}$ (i.e. to a different set of doubly-occupied orbitals), and each one partially separable with respect to a given partition. If some of the contributions in the RDO include only singly-occupied orbitals, then the $n$BRDM can in principle display genuine multipartite entanglement. 
Third, a term of the RDO like the one reported in Eq.~\eqref{xx1} does not contribute to the $n$BRDM if the state $| \psi_{\overline {1\dots p}}\rangle$ is replaced by one where at least one of the detection modes $\phi_i$ is unoccupied. As commented below, this can be exploited to remove undesired contributions from the $n$BRDM.

Overall, the double occupations of the detection modes are detrimental to {\color{black} the extraction of} genuine multipartite entanglement. One way to suppress the effect of these contributions on the $n$BRDM is to consider fermionic states $|\Psi\rangle$ with exactly $n$ electrons in a given subspace, spanned by the detection orbitals $\phi_i$ \cite{Troiani25a}.

\subsection{The effect of rotational invariance}

The projection of the RDOs on the $S=0$ subspace corresponds to a {\color{black}rotationally invariant} pure state. However, {\color{black}in order to be rotationally invariant,} the projection on a subspace $S>0$ is necessarily given by a mixture of terms belonging to subspaces with different values of $M$ (see Appendix \ref{appendixA}). This tends to decrease the amount of entanglement, as shown below through the use of spin-squeezing inequalities. {\color{black} A fully separable state on $n$ spins necessarily fulfills the following inequalities} \cite{Toth09a}:
\begin{align}
(\Delta S_x)^2 + (\Delta S_y)^2 + (\Delta S_z)^2 &\ge \frac{n}{2} \\
\langle S_\alpha^2 \rangle + \langle S_\beta^2 \rangle - \frac{n}{2} &\le (n-1) (\Delta S_\gamma)^2 \\
(n-1) [(\Delta S_\alpha)^2 + (\Delta S_\beta)^2] &\ge \langle S_\gamma^2 \rangle + \frac{1}{4} n(n-2)\,.
\end{align}
{\color{black} The violation of one (or more) of the above inequalities thus implies the presence of some form of entanglement.} 

For the eigenstates of the total spin and of its projection along the quantization axis, $|S,M\rangle$, one has that:
$ \langle S_x^2 \rangle \!=\! \langle S_y^2 \rangle \!=\! (\Delta S_x)^2 \!=\! (\Delta S_y)^2 \!=\! \frac{1}{2} [S(S\!+\!1)\!-\!M^2]$, 
$\langle S_z^2 \rangle = M^2$, and $(\Delta S_z)^2 = 0 $.
From this, it follows that:
\begin{align}
    S(S+1)-M^2 &\ge \frac{n}{2} \\
    S(S+1)-M^2 &\le \frac{n}{2} \\
(N-1) [S(S+1)-M^2] &\ge M^2 + \frac{1}{4} n(n-2) \,,
\end{align}
where, in the last two inequalities, we consider the case $\gamma = z$. The first inequality tends to be violated by states with low values of $S$ and relatively large $|M|$. 
The second inequality tends to be violated by states with high values of $S$ and small values of $|M|$. 
The only states $|S,M\rangle$ that satisfy both the first and second inequalities are those with $S=|M|=n/2$, which in fact are the only fully separable ones. The third inequality above tends to be violated by states with low values of $S$ and large values of $|M|$.

Rotational invariance within a subspace of given $S$ can only be achieved by the mixtures: $\hat\rho = \frac{1}{2S+1} \sum_{M=-S}^S |M \rangle\langle M|$, where we omit for simplicity other quantum numbers (see Appendix \ref{appendixA} for a discussion of this point). In this case, one has that:
$\langle S_x^2 \rangle = \langle S_y^2 \rangle = \langle S_z^2 \rangle = (\Delta S_x)^2 = (\Delta S_y)^2 = (\Delta S_z)^2 = \frac{1}{3} S(S+1)$.
From this and from the equation $\sum_{M=-S}^S M^2 = \frac{S}{3}(S+1)(2S+1)$ it follows that:
\begin{align}
   \langle {\bf S}^2 \rangle &\ge \frac{n}{2} \label{ineq} \\
    - \frac{3n}{2} &\le \langle {\bf S}^2 \rangle (n-3) \\
    \langle {\bf S}^2 \rangle &\ge \frac{3n(n-2)}{4(2n-3)} \,, 
\end{align}
where $\langle {\bf S}^2 \rangle = S(S+1)$.
The first inequality can be violated by low-spin states, for example by $S=0$ for $n=4$, or $S\le 1$ for $n=6$. The second inequality is always (i.e. for any value of $n$ and $S\le n/2$) satisfied, and therefore cannot be used to detect entanglement in the rotational invariant density operators.
The third inequality is less stringent than the first one, and can also be violated by low-spin states, for example by $S=0$ for $n=4,6$. It will thus suffice to consider the first inequality in order to detect entanglement in rotationally invariant spin states.

Let's consider a RDO with $n$ singly occupied modes and corresponding to a singlet state. The {\color{black} resulting} $n$BRDM also corresponds to a spin singlet, and thus violates the inequality in Eq.~\eqref{ineq}, being $\langle {\bf S}^2 \rangle = 0$. In the presence of $l$ doubly occupied orbitals $\phi_i$ in the singlet RDO, the $n$BRDM is given by the product of a singlet state formed by the $(n-l)$ (assumed even) spins detected in the singly occupied modes, and of an identity operator for the remaining $l$ spins. These determine the value of $\langle {\bf S}^2 \rangle$, which is $3/4, 3/2, 2, 3, \dots$ for $l=1,2,3,4,\dots$; the spin state is thus (detected to be) entangled for $n$ larger than $3, 4,5,8,\dots$
These relations show that, generally speaking, the detection of entanglement in rotationally invariant states is related to low expectation values of the total spin: the lower the {\color{black}value of} $n$, the lower the threshold value for $\langle {\bf S}^2 \rangle$. This condition calls for a limited value of doubly occupied orbitals in the relevant RDO. }

\section{Conclusions\label{sec6}}

In this work, toward the analysis of entanglement for the case of indistinguishable particles, we have investigated the structure of the spin states that can be extracted from closed-shell states of $N$ fermions. First, we have shown that a change in the electron modes (orbitals), used as labels for the spins, gives rise to a state with multiple configurations; each of these is given by the product of $N/2$ singlets, either localized in doubly-occupied orbitals or delocalized over two different singly-occupied orbitals. Second, we have derived the structure of the reduced density operators, defining the relevant state of $n$-mode subsystems. 
In particular, we have identified the subspaces that block-diagonalize the reduced density operators for $n\le 6$.  
Third, we have derived the $n$-body reduced density matrices from the relevant $n$-mode reduced density operators. 
Finally, we have shown that both the rotational invariance and  the presence of double occupations in the detection {\color{black}modes limit the amount} of spin entanglement, which can {\color{black}nonetheless be extracted} for sufficiently low expectation values of the total spin operator.

{\color{black} The extraction of the $n$ particles from a state that differs from a closed shell $N$-electron state implies different constraints for the RDOs and the $n$BRDOs. In particular, a pure fermionic state $|\Psi\rangle$ characterized by an $S>0$, or including components with nonzero values of $S$, cannot be characterized by rotational invariance in the spin space. This implies that the $n$BRDMs need not correspond to singlet spin states, or to mixtures, with equal weights, of components with equal values of $S$ and different values of $M$. Fermionic states $|\Psi\rangle$ characterized also by an undefined value of the total spin projection might give rise to spin states with coherences between different values of $M$. Finally, the replacement of a pure fermionic state $|\Psi\rangle$ with a rotationally invariant mixture of different terms, each one necessarily characterized by a defined value of the particle number (due to the superselection rules) and of the total spin (due to rotational invariance), would give rise to RDOs and $n$BRDOs with the same formal properties as those discussed in the present analysis. We finally note that removing of the constraints related to a singlet fermionic state $|\Psi\rangle$ might in principle allow the extraction of other forms of entangled spin states, beyond the ones discussed in Ref. \cite{Troiani25a} and in the present paper.}

\acknowledgments

The authors acknowledge stimulating discussions with Celestino Angeli, and financial support from the Ministero dell’Universit\`a e della Ricerca (MUR) under the Project PRIN 2022 number 2022W9W423. F. T. and A. S.  acknowledge financial support from  the Ministero dell’Universit\`a e della Ricerca (MUR) under  PNRR Project PE0000023-NQSTI.  

\appendix

\section{General properties of reduced density operators\label{appendixA}}

Here, we illustrate the restrictions on the reduced density operators (RDOs), which follow from the properties of close-shell $N$-particle state we analyzed in Sec. \ref{sec2}. For obtaining the RDOs that define the state of $K$ modes, we must perform a partial trace of $\hat{\rho} = |\Psi\rangle\langle\Psi|$  over the complementary modes. Essentially, we  take advantage of  the (quasi-)tensor product of the Fock states.

\subsubsection{Particle number}

The reduced density matrix $\hat{\rho}_\mathcal{S}$ for subsystem $\mathcal{S}$ generally consists of terms characterized by different particle numbers, but cannot include cross terms $|{\bf V}\rangle \langle {\bf V}'|$ ($|{\bf V}\rangle$ and $|{\bf V}'\rangle$ being two generic $N$-electron configurations) where $N_{\bf V} \neq N_{{\bf V}'}$. This results from the fact that: $(i)$ $|\Psi\rangle$ and (thus) the original density operator $\hat{\rho}\,{\color{black} = |\Psi\rangle\langle\Psi|}$ are characterized by a defined number of electrons $N$; $(ii)$ the particle number is a local operator with respect to the present partition in modes, being 
$\hat{N} = \sum_k \hat{n}_k$, with $\hat{n}_k = \hat{d}^\dagger_{k,\uparrow} \hat{d}_{k,\uparrow} + \hat{d}^\dagger_{k,\downarrow} \hat{d}_{k,\downarrow} $  the number operator relative to mode $\phi_k$ (see Appendix \ref{appendixB} for further details). 

\subsubsection{Total spin projection}

The reduced density matrix $\hat{\rho}_\mathcal{S}$ for subsystem $\mathcal{S}$ generally consists of terms characterized by different values of the total spin projection $M_{\bf V}$, but cannot include coherences between states $|{\bf V}\rangle$ and $|{\bf V}'\rangle$ with $M_{\bf V} \neq M_{{\bf V}'}$. This results from the fact that: $(i)$ $|\Psi\rangle$ and (thus) the original density operator $\hat{\rho}\,{\color{black} = |\Psi\rangle\langle\Psi|}$ are characterized by a defined value of the total spin projection ($M=0$); $(ii)$ the total spin projection number is a local operator with respect to the present partition in modes, being 
$\hat{S}_z =   \sum_k  \hat{s}_{z,k} $, where $\hat{s}_{z,k}= \frac{1}{2} (\hat{d}^\dagger_{k,\uparrow} \hat{d}_{k,\uparrow} - \hat{d}^\dagger_{k,\downarrow} \hat{d}_{k,\downarrow})$ the spin-projection operator relative to the orbital $\phi_k$ (see Appendix \ref{appendixB} for further details). 

\subsubsection{Rotational invariance}

The original state $|\Psi\rangle$ is given by the product of spin singlets, and is thus invariant with respect to arbitrary rotations in the spin space. The RDOs in general include, besides singlet states, variable numbers of unpaired spins. However, they retain the rotational invariance that characterizes $\hat{\rho} = |\Psi\rangle\langle\Psi|$. This has two fundamental implications: 
$(i)$ the RDOs cannot include off-diagonal terms between states corresponding to different values of the total spin $S$ \cite{Breuer05a}; $(ii)$ the terms that belong to identical values of all the quantum numbers but the total spin projection must have the same expression. In other words, let us consider the generic contribution in the $n$-mode RDO:
\begin{align}
    \hat{\rho}^{(n)}_{S,M;{\bf n}} = \sum_{\gamma,\gamma'} \rho_{\gamma,\gamma'}^{(S,M)} |S,M;\gamma\rangle\langle S,M;\gamma'|\,,
\end{align}
where $\gamma,\gamma'$ define all the quantum numbers that can vary within the subspace (such as the partial spin sums). The above property implies that the matrix elements $\rho_{\gamma,\gamma'}^{(S,M)}$ are the same for all values of $M$ (with $-S\le M \le S$).

\section{Allowed coherences in the reduced density operators\label{appendixB}}

Let us consider a bipartite system, formed by the product of two subsystems, $\mathcal{A}$ and $\mathcal{B}$, and an observable $\hat{X}$ that is local with respect to such bipartition:
\begin{align}
    \hat{X} = \hat{X}_{\mathcal{A}} \otimes \hat{I}_{\mathcal{B}} + \hat{I}_{\mathcal{A}} \otimes \hat{X}_{\mathcal{B}}\,, 
\end{align}
$\hat{I}$ being the identity operator.
We assume that the state $|\Psi\rangle$ of the overall system is an eigenstate of the observable $\hat{X}$: $\hat{X} |\Psi\rangle = X |\Psi\rangle$. 
The system state can always be expanded in the form:
\begin{align}
    |\Psi\rangle = \sum_{X_\mathcal{A}} C({X_\mathcal{A}}) | \psi_\mathcal{A} (X_\mathcal{A}) , \psi_{\mathcal{B}} (X-X_\mathcal{A}) \rangle \,,
\end{align}
$ | \psi_\mathcal{A} (X_\mathcal{A}) \rangle $ and $| \psi_{\mathcal{B}} (X-X_\mathcal{A}) \rangle$ being the eigenstates of $\hat{X}_{\mathcal{A}}$ and  $\hat{X}_{\mathcal{B}}$, respectively:
\begin{align}
    \hat{X}_{\mathcal{A}} \, | \psi_\mathcal{A} (X_\mathcal{A}) \rangle = X_\mathcal{A}\,| \psi_\mathcal{A} (X_\mathcal{A}) \rangle \,, \\
    \hat{X}_{\mathcal{B}} \, | \psi_{\mathcal{B}} (X_{\mathcal{B}}) \rangle = X_{\mathcal{B}}\,| \psi_{\mathcal{B}} (X_\mathcal{A}) \rangle\,.
\end{align}
In order to derive the reduced density operator of the subsystem $\mathcal{A}$, one performs the partial trace on $\mathcal{B}$. Since the states $| \psi_{\mathcal{B}} (X_{\mathcal{B}}) \rangle$ and $| \psi_{\mathcal{B}} (X_{\mathcal{B}}') \rangle$ are orthogonal for any $X_{\mathcal{B}}\neq X_{\mathcal{B}}'$, this gives:
\begin{align}
    \hat{\rho}_\mathcal{A} = {\rm Tr}_{\mathcal{B}} (|\Psi\rangle\langle\Psi|) = \sum_{X_\mathcal{A}} | C({X_\mathcal{A}}) |^2 |\psi_\mathcal{A} (X_\mathcal{A})\rangle\langle\psi_\mathcal{A} (X_\mathcal{A})|\,,
\end{align}
which contains no coherences (off-diagonal terms) between states with different values of $X_\mathcal{A}$. 

\section{Overcomplete bases in few-spin systems}\label{appendixC}

Hereafter, we show that real density matrices of three- and four-spin systems can be written as combination of projectors on states $|{\bf V}\rangle$. As in the rest of the paper, we adopt the following convention: $S_{ij}$ stands for a spin singlet ($S_{ij}=0$), while the triplet is denoted by $S_{ij}=1$.

Three 1/2 spins form a two-dimensional $S=M=1/2$ subspace, spanned by the states: 
\begin{align}
|S_{12},\uparrow_3\rangle & = \frac{1}{\sqrt{2}}(|\!\uparrow_1,\downarrow_2,\uparrow_3\rangle - |\!\downarrow_1,\uparrow_2,\uparrow_3\rangle) \,, \\
|S_{12}\!=\!1,\uparrow_3\rangle \!&=\! \frac{1}{2} (|\!\uparrow_1,\downarrow_2,\uparrow_3\rangle \!+\! |\!\downarrow_1,\uparrow_2,\uparrow_3\rangle \!-\! 2 |\!\uparrow_1,\uparrow_2,\downarrow_3\rangle) \,.
\end{align}
The other states we refer to in the text are given by:
\begin{align}
|S_{23},\uparrow_1\rangle = \frac{1}{\sqrt{2}}(|\uparrow_1,\uparrow_2,\downarrow_3\rangle - |\uparrow_1,\downarrow_2,\uparrow_3\rangle) \,, \\
|S_{13},\uparrow_2\rangle = \frac{1}{\sqrt{2}}(|\uparrow_1,\uparrow_2,\downarrow_3\rangle - |\downarrow_1,\uparrow_2,\uparrow_3\rangle) \,.
\end{align}
The matricial expressions of the projectors on the three singlet states above, in the basis $\{|S_{12},\uparrow_3\rangle,|S_{12}=1,\uparrow_3\rangle\}$, are given by:
\begin{align}
    |S_{12},\uparrow_3 \rangle\langle S_{12},\uparrow_{3}| &= \left( \begin{array}{cc}
        1 & 0 \\
        0 & 0
    \end{array}\right) \,, \\
    |S_{13},\uparrow_{2} \rangle\langle S_{13},\uparrow_{2}| &= \frac{1}{4} \left( \begin{array}{cc}
        1 & -\sqrt{3} \\
        -\sqrt{3} & 3
    \end{array}\right) \,, \\
    |S_{23},\uparrow_{1} \rangle\langle S_{23},\uparrow_{1}| &= \frac{1}{4} \left( \begin{array}{cc}
        1 & \sqrt{3} \\
        \sqrt{3} & 3
    \end{array}\right) \,.
\end{align}
One can thus decompose any density matrix with (three independent) real elements into a linear combination with real coefficients of the above (linearly independent) matrices. It should be stressed, however, that these coefficients need not be all positive. Therefore, the above decomposition might not physically correspond to a statistical mixture of states.

The case of four spins is discussed in the Supplemental Material of Ref. \cite{Troiani25a}.

}


\begin{thebibliography}{45}%
\makeatletter
\providecommand \@ifxundefined [1]{%
 \@ifx{#1\undefined}
}%
\providecommand \@ifnum [1]{%
 \ifnum #1\expandafter \@firstoftwo
 \else \expandafter \@secondoftwo
 \fi
}%
\providecommand \@ifx [1]{%
 \ifx #1\expandafter \@firstoftwo
 \else \expandafter \@secondoftwo
 \fi
}%
\providecommand \natexlab [1]{#1}%
\providecommand \enquote  [1]{``#1''}%
\providecommand \bibnamefont  [1]{#1}%
\providecommand \bibfnamefont [1]{#1}%
\providecommand \citenamefont [1]{#1}%
\providecommand \href@noop [0]{\@secondoftwo}%
\providecommand \href [0]{\begingroup \@sanitize@url \@href}%
\providecommand \@href[1]{\@@startlink{#1}\@@href}%
\providecommand \@@href[1]{\endgroup#1\@@endlink}%
\providecommand \@sanitize@url [0]{\catcode `\\12\catcode `\$12\catcode `\&12\catcode `\#12\catcode `\^12\catcode `\_12\catcode `\%12\relax}%
\providecommand \@@startlink[1]{}%
\providecommand \@@endlink[0]{}%
\providecommand \url  [0]{\begingroup\@sanitize@url \@url }%
\providecommand \@url [1]{\endgroup\@href {#1}{\urlprefix }}%
\providecommand \urlprefix  [0]{URL }%
\providecommand \Eprint [0]{\href }%
\providecommand \doibase [0]{http://dx.doi.org/}%
\providecommand \selectlanguage [0]{\@gobble}%
\providecommand \bibinfo  [0]{\@secondoftwo}%
\providecommand \bibfield  [0]{\@secondoftwo}%
\providecommand \translation [1]{[#1]}%
\providecommand \BibitemOpen [0]{}%
\providecommand \bibitemStop [0]{}%
\providecommand \bibitemNoStop [0]{.\EOS\space}%
\providecommand \EOS [0]{\spacefactor3000\relax}%
\providecommand \BibitemShut  [1]{\csname bibitem#1\endcsname}%
\let\auto@bib@innerbib\@empty
\bibitem [{\citenamefont {Nielsen}\ and\ \citenamefont {Chuang}(2010)}]{Nielsen_Chuang_2010}%
  \BibitemOpen
  \bibfield  {author} {\bibinfo {author} {\bibfnamefont {M.~A.}\ \bibnamefont {Nielsen}}\ and\ \bibinfo {author} {\bibfnamefont {I.~L.}\ \bibnamefont {Chuang}},\ }\href@noop {} {\emph {\bibinfo {title} {Quantum Computation and Quantum Information: 10th Anniversary Edition}}}\ (\bibinfo  {publisher} {Cambridge University Press},\ \bibinfo {year} {2010})\BibitemShut {NoStop}%
\bibitem [{\citenamefont {Horodecki}\ \emph {et~al.}(2009)\citenamefont {Horodecki}, \citenamefont {Horodecki}, \citenamefont {Horodecki},\ and\ \citenamefont {Horodecki}}]{Horodecki09a}%
  \BibitemOpen
  \bibfield  {author} {\bibinfo {author} {\bibfnamefont {R.}~\bibnamefont {Horodecki}}, \bibinfo {author} {\bibfnamefont {P.}~\bibnamefont {Horodecki}}, \bibinfo {author} {\bibfnamefont {M.}~\bibnamefont {Horodecki}}, \ and\ \bibinfo {author} {\bibfnamefont {K.}~\bibnamefont {Horodecki}},\ }\href {\doibase 10.1103/RevModPhys.81.865} {\bibfield  {journal} {\bibinfo  {journal} {Rev. Mod. Phys.}\ }\textbf {\bibinfo {volume} {81}},\ \bibinfo {pages} {865} (\bibinfo {year} {2009})}\BibitemShut {NoStop}%
\bibitem [{\citenamefont {Benatti}\ \emph {et~al.}(2020)\citenamefont {Benatti}, \citenamefont {Floreanini}, \citenamefont {Franchini},\ and\ \citenamefont {Marzolino}}]{Benatti20a}%
  \BibitemOpen
  \bibfield  {author} {\bibinfo {author} {\bibfnamefont {F.}~\bibnamefont {Benatti}}, \bibinfo {author} {\bibfnamefont {R.}~\bibnamefont {Floreanini}}, \bibinfo {author} {\bibfnamefont {F.}~\bibnamefont {Franchini}}, \ and\ \bibinfo {author} {\bibfnamefont {U.}~\bibnamefont {Marzolino}},\ }\href {\doibase https://doi.org/10.1016/j.physrep.2020.07.003} {\bibfield  {journal} {\bibinfo  {journal} {Physics Reports}\ }\textbf {\bibinfo {volume} {878}},\ \bibinfo {pages} {1} (\bibinfo {year} {2020})},\ \bibinfo {note} {entanglement in indistinguishable particle systems}\BibitemShut {NoStop}%
\bibitem [{\citenamefont {Catren}(2023)}]{PNASA2023}%
  \BibitemOpen
  \bibfield  {author} {\bibinfo {author} {\bibfnamefont {G.}~\bibnamefont {Catren}},\ }\href {\doibase 10.1098/rsta.2022.0109} {\bibfield  {journal} {\bibinfo  {journal} {Philosophical Transactions of the Royal Society A: Mathematical, Physical and Engineering Sciences}\ }\textbf {\bibinfo {volume} {381}},\ \bibinfo {pages} {20220109} (\bibinfo {year} {2023})}\BibitemShut {NoStop}%
\bibitem [{\citenamefont {Schliemann}\ \emph {et~al.}(2001)\citenamefont {Schliemann}, \citenamefont {Cirac}, \citenamefont {Ku\ifmmode~\acute{s}\else \'{s}\fi{}}, \citenamefont {Lewenstein},\ and\ \citenamefont {Loss}}]{Schliemann01a}%
  \BibitemOpen
  \bibfield  {author} {\bibinfo {author} {\bibfnamefont {J.}~\bibnamefont {Schliemann}}, \bibinfo {author} {\bibfnamefont {J.~I.}\ \bibnamefont {Cirac}}, \bibinfo {author} {\bibfnamefont {M.}~\bibnamefont {Ku\ifmmode~\acute{s}\else \'{s}\fi{}}}, \bibinfo {author} {\bibfnamefont {M.}~\bibnamefont {Lewenstein}}, \ and\ \bibinfo {author} {\bibfnamefont {D.}~\bibnamefont {Loss}},\ }\href {\doibase 10.1103/PhysRevA.64.022303} {\bibfield  {journal} {\bibinfo  {journal} {Phys. Rev. A}\ }\textbf {\bibinfo {volume} {64}},\ \bibinfo {pages} {022303} (\bibinfo {year} {2001})}\BibitemShut {NoStop}%
\bibitem [{\citenamefont {Li}\ \emph {et~al.}(2001)\citenamefont {Li}, \citenamefont {Zeng}, \citenamefont {Liu},\ and\ \citenamefont {Long}}]{Li01a}%
  \BibitemOpen
  \bibfield  {author} {\bibinfo {author} {\bibfnamefont {Y.~S.}\ \bibnamefont {Li}}, \bibinfo {author} {\bibfnamefont {B.}~\bibnamefont {Zeng}}, \bibinfo {author} {\bibfnamefont {X.~S.}\ \bibnamefont {Liu}}, \ and\ \bibinfo {author} {\bibfnamefont {G.~L.}\ \bibnamefont {Long}},\ }\href {\doibase 10.1103/PhysRevA.64.054302} {\bibfield  {journal} {\bibinfo  {journal} {Phys. Rev. A}\ }\textbf {\bibinfo {volume} {64}},\ \bibinfo {pages} {054302} (\bibinfo {year} {2001})}\BibitemShut {NoStop}%
\bibitem [{\citenamefont {Zanardi}(2002)}]{Zanardi02a}%
  \BibitemOpen
  \bibfield  {author} {\bibinfo {author} {\bibfnamefont {P.}~\bibnamefont {Zanardi}},\ }\href {\doibase 10.1103/PhysRevA.65.042101} {\bibfield  {journal} {\bibinfo  {journal} {Phys. Rev. A}\ }\textbf {\bibinfo {volume} {65}},\ \bibinfo {pages} {042101} (\bibinfo {year} {2002})}\BibitemShut {NoStop}%
\bibitem [{\citenamefont {Amico}\ \emph {et~al.}(2008)\citenamefont {Amico}, \citenamefont {Fazio}, \citenamefont {Osterloh},\ and\ \citenamefont {Vedral}}]{Amico08a}%
  \BibitemOpen
  \bibfield  {author} {\bibinfo {author} {\bibfnamefont {L.}~\bibnamefont {Amico}}, \bibinfo {author} {\bibfnamefont {R.}~\bibnamefont {Fazio}}, \bibinfo {author} {\bibfnamefont {A.}~\bibnamefont {Osterloh}}, \ and\ \bibinfo {author} {\bibfnamefont {V.}~\bibnamefont {Vedral}},\ }\href {\doibase 10.1103/RevModPhys.80.517} {\bibfield  {journal} {\bibinfo  {journal} {Rev. Mod. Phys.}\ }\textbf {\bibinfo {volume} {80}},\ \bibinfo {pages} {517} (\bibinfo {year} {2008})}\BibitemShut {NoStop}%
\bibitem [{\citenamefont {Fran\ifmmode~\mbox{\c{c}}\else \c{c}\fi{}a}\ and\ \citenamefont {Capelle}(2008)}]{Franca08a}%
  \BibitemOpen
  \bibfield  {author} {\bibinfo {author} {\bibfnamefont {V.~V.}\ \bibnamefont {Fran\ifmmode~\mbox{\c{c}}\else \c{c}\fi{}a}}\ and\ \bibinfo {author} {\bibfnamefont {K.}~\bibnamefont {Capelle}},\ }\href {\doibase 10.1103/PhysRevLett.100.070403} {\bibfield  {journal} {\bibinfo  {journal} {Phys. Rev. Lett.}\ }\textbf {\bibinfo {volume} {100}},\ \bibinfo {pages} {070403} (\bibinfo {year} {2008})}\BibitemShut {NoStop}%
\bibitem [{\citenamefont {Gigena}\ and\ \citenamefont {Rossignoli}(2015)}]{Gigena15a}%
  \BibitemOpen
  \bibfield  {author} {\bibinfo {author} {\bibfnamefont {N.}~\bibnamefont {Gigena}}\ and\ \bibinfo {author} {\bibfnamefont {R.}~\bibnamefont {Rossignoli}},\ }\href {\doibase 10.1103/PhysRevA.92.042326} {\bibfield  {journal} {\bibinfo  {journal} {Phys. Rev. A}\ }\textbf {\bibinfo {volume} {92}},\ \bibinfo {pages} {042326} (\bibinfo {year} {2015})}\BibitemShut {NoStop}%
\bibitem [{\citenamefont {Fran\ifmmode~\mbox{\c{c}}\else \c{c}\fi{}a}\ and\ \citenamefont {D'Amico}(2011)}]{Franca11a}%
  \BibitemOpen
  \bibfield  {author} {\bibinfo {author} {\bibfnamefont {V.~V.}\ \bibnamefont {Fran\ifmmode~\mbox{\c{c}}\else \c{c}\fi{}a}}\ and\ \bibinfo {author} {\bibfnamefont {I.}~\bibnamefont {D'Amico}},\ }\href {\doibase 10.1103/PhysRevA.83.042311} {\bibfield  {journal} {\bibinfo  {journal} {Phys. Rev. A}\ }\textbf {\bibinfo {volume} {83}},\ \bibinfo {pages} {042311} (\bibinfo {year} {2011})}\BibitemShut {NoStop}%
\bibitem [{\citenamefont {Legeza}\ and\ \citenamefont {S\'olyom}(2003)}]{Legeza2003}%
  \BibitemOpen
  \bibfield  {author} {\bibinfo {author} {\bibfnamefont {O.}~\bibnamefont {Legeza}}\ and\ \bibinfo {author} {\bibfnamefont {J.}~\bibnamefont {S\'olyom}},\ }\href {\doibase 10.1103/PhysRevB.68.195116} {\bibfield  {journal} {\bibinfo  {journal} {Phys. Rev. B}\ }\textbf {\bibinfo {volume} {68}},\ \bibinfo {pages} {195116} (\bibinfo {year} {2003})}\BibitemShut {NoStop}%
\bibitem [{\citenamefont {Rissler}\ \emph {et~al.}(2006)\citenamefont {Rissler}, \citenamefont {Noack},\ and\ \citenamefont {White}}]{Rissler2006}%
  \BibitemOpen
  \bibfield  {author} {\bibinfo {author} {\bibfnamefont {J.}~\bibnamefont {Rissler}}, \bibinfo {author} {\bibfnamefont {R.~M.}\ \bibnamefont {Noack}}, \ and\ \bibinfo {author} {\bibfnamefont {S.~R.}\ \bibnamefont {White}},\ }\href {\doibase https://doi.org/10.1016/j.chemphys.2005.10.018} {\bibfield  {journal} {\bibinfo  {journal} {Chemical Physics}\ }\textbf {\bibinfo {volume} {323}},\ \bibinfo {pages} {519} (\bibinfo {year} {2006})}\BibitemShut {NoStop}%
\bibitem [{\citenamefont {Boguslawski}\ and\ \citenamefont {Tecmer}(2015)}]{Boguslawski2015}%
  \BibitemOpen
  \bibfield  {author} {\bibinfo {author} {\bibfnamefont {K.}~\bibnamefont {Boguslawski}}\ and\ \bibinfo {author} {\bibfnamefont {P.}~\bibnamefont {Tecmer}},\ }\href {\doibase https://doi.org/10.1002/qua.24832} {\bibfield  {journal} {\bibinfo  {journal} {International Journal of Quantum Chemistry}\ }\textbf {\bibinfo {volume} {115}},\ \bibinfo {pages} {1289} (\bibinfo {year} {2015})}\BibitemShut {NoStop}%
\bibitem [{\citenamefont {Boguslawski}\ \emph {et~al.}(2013)\citenamefont {Boguslawski}, \citenamefont {Tecmer}, \citenamefont {Barcza}, \citenamefont {Legeza},\ and\ \citenamefont {Reiher}}]{Boguslawski2013}%
  \BibitemOpen
  \bibfield  {author} {\bibinfo {author} {\bibfnamefont {K.}~\bibnamefont {Boguslawski}}, \bibinfo {author} {\bibfnamefont {P.}~\bibnamefont {Tecmer}}, \bibinfo {author} {\bibfnamefont {G.}~\bibnamefont {Barcza}}, \bibinfo {author} {\bibfnamefont {O.}~\bibnamefont {Legeza}}, \ and\ \bibinfo {author} {\bibfnamefont {M.}~\bibnamefont {Reiher}},\ }\href {\doibase 10.1021/ct400247p} {\bibfield  {journal} {\bibinfo  {journal} {Journal of Chemical Theory and Computation}\ }\textbf {\bibinfo {volume} {9}},\ \bibinfo {pages} {2959} (\bibinfo {year} {2013})},\ \bibinfo {note} {pMID: 26583979}\BibitemShut {NoStop}%
\bibitem [{\citenamefont {Duperrouzel}\ \emph {et~al.}(2015)\citenamefont {Duperrouzel}, \citenamefont {Tecmer}, \citenamefont {Boguslawski}, \citenamefont {Barcza}, \citenamefont {Örs Legeza},\ and\ \citenamefont {Ayers}}]{Duperrouzel2015}%
  \BibitemOpen
  \bibfield  {author} {\bibinfo {author} {\bibfnamefont {C.}~\bibnamefont {Duperrouzel}}, \bibinfo {author} {\bibfnamefont {P.}~\bibnamefont {Tecmer}}, \bibinfo {author} {\bibfnamefont {K.}~\bibnamefont {Boguslawski}}, \bibinfo {author} {\bibfnamefont {G.}~\bibnamefont {Barcza}}, \bibinfo {author} {\bibnamefont {Örs Legeza}}, \ and\ \bibinfo {author} {\bibfnamefont {P.~W.}\ \bibnamefont {Ayers}},\ }\href {\doibase https://doi.org/10.1016/j.cplett.2015.01.005} {\bibfield  {journal} {\bibinfo  {journal} {Chemical Physics Letters}\ }\textbf {\bibinfo {volume} {621}},\ \bibinfo {pages} {160} (\bibinfo {year} {2015})}\BibitemShut {NoStop}%
\bibitem [{\citenamefont {Stein}\ and\ \citenamefont {Reiher}(2016)}]{Stein2016}%
  \BibitemOpen
  \bibfield  {author} {\bibinfo {author} {\bibfnamefont {C.~J.}\ \bibnamefont {Stein}}\ and\ \bibinfo {author} {\bibfnamefont {M.}~\bibnamefont {Reiher}},\ }\href {\doibase 10.1021/acs.jctc.6b00156} {\bibfield  {journal} {\bibinfo  {journal} {Journal of Chemical Theory and Computation}\ }\textbf {\bibinfo {volume} {12}},\ \bibinfo {pages} {1760} (\bibinfo {year} {2016})},\ \bibinfo {note} {pMID: 26959891}\BibitemShut {NoStop}%
\bibitem [{\citenamefont {Ding}\ \emph {et~al.}(2021)\citenamefont {Ding}, \citenamefont {Mardazad}, \citenamefont {Das}, \citenamefont {Szalay}, \citenamefont {Schollwöck}, \citenamefont {Zimborás},\ and\ \citenamefont {Schilling}}]{Ding2021}%
  \BibitemOpen
  \bibfield  {author} {\bibinfo {author} {\bibfnamefont {L.}~\bibnamefont {Ding}}, \bibinfo {author} {\bibfnamefont {S.}~\bibnamefont {Mardazad}}, \bibinfo {author} {\bibfnamefont {S.}~\bibnamefont {Das}}, \bibinfo {author} {\bibfnamefont {S.}~\bibnamefont {Szalay}}, \bibinfo {author} {\bibfnamefont {U.}~\bibnamefont {Schollwöck}}, \bibinfo {author} {\bibfnamefont {Z.}~\bibnamefont {Zimborás}}, \ and\ \bibinfo {author} {\bibfnamefont {C.}~\bibnamefont {Schilling}},\ }\href {\doibase 10.1021/acs.jctc.0c00559} {\bibfield  {journal} {\bibinfo  {journal} {Journal of Chemical Theory and Computation}\ }\textbf {\bibinfo {volume} {17}},\ \bibinfo {pages} {79} (\bibinfo {year} {2021})},\ \bibinfo {note} {pMID: 33430597}\BibitemShut {NoStop}%
\bibitem [{\citenamefont {Wiseman}\ and\ \citenamefont {Vaccaro}(2003)}]{Wiseman03a}%
  \BibitemOpen
  \bibfield  {author} {\bibinfo {author} {\bibfnamefont {H.~M.}\ \bibnamefont {Wiseman}}\ and\ \bibinfo {author} {\bibfnamefont {J.~A.}\ \bibnamefont {Vaccaro}},\ }\href {\doibase 10.1103/PhysRevLett.91.097902} {\bibfield  {journal} {\bibinfo  {journal} {Phys. Rev. Lett.}\ }\textbf {\bibinfo {volume} {91}},\ \bibinfo {pages} {097902} (\bibinfo {year} {2003})}\BibitemShut {NoStop}%
\bibitem [{\citenamefont {Bartlett}\ and\ \citenamefont {Wiseman}(2003)}]{Bartlett03a}%
  \BibitemOpen
  \bibfield  {author} {\bibinfo {author} {\bibfnamefont {S.~D.}\ \bibnamefont {Bartlett}}\ and\ \bibinfo {author} {\bibfnamefont {H.~M.}\ \bibnamefont {Wiseman}},\ }\href {\doibase 10.1103/PhysRevLett.91.097903} {\bibfield  {journal} {\bibinfo  {journal} {Phys. Rev. Lett.}\ }\textbf {\bibinfo {volume} {91}},\ \bibinfo {pages} {097903} (\bibinfo {year} {2003})}\BibitemShut {NoStop}%
\bibitem [{\citenamefont {Friis}\ \emph {et~al.}(2014)\citenamefont {Friis}, \citenamefont {Dunjko}, \citenamefont {D\"ur},\ and\ \citenamefont {Briegel}}]{Friis2014}%
  \BibitemOpen
  \bibfield  {author} {\bibinfo {author} {\bibfnamefont {N.}~\bibnamefont {Friis}}, \bibinfo {author} {\bibfnamefont {V.}~\bibnamefont {Dunjko}}, \bibinfo {author} {\bibfnamefont {W.}~\bibnamefont {D\"ur}}, \ and\ \bibinfo {author} {\bibfnamefont {H.~J.}\ \bibnamefont {Briegel}},\ }\href {\doibase 10.1103/PhysRevA.89.030303} {\bibfield  {journal} {\bibinfo  {journal} {Phys. Rev. A}\ }\textbf {\bibinfo {volume} {89}},\ \bibinfo {pages} {030303} (\bibinfo {year} {2014})}\BibitemShut {NoStop}%
\bibitem [{\citenamefont {Friis}(2016)}]{Friis2016}%
  \BibitemOpen
  \bibfield  {author} {\bibinfo {author} {\bibfnamefont {N.}~\bibnamefont {Friis}},\ }\href {\doibase 10.1088/1367-2630/18/3/033014} {\bibfield  {journal} {\bibinfo  {journal} {New Journal of Physics}\ }\textbf {\bibinfo {volume} {18}},\ \bibinfo {pages} {033014} (\bibinfo {year} {2016})}\BibitemShut {NoStop}%
\bibitem [{\citenamefont {Killoran}\ \emph {et~al.}(2014)\citenamefont {Killoran}, \citenamefont {Cramer},\ and\ \citenamefont {Plenio}}]{Killoran14a}%
  \BibitemOpen
  \bibfield  {author} {\bibinfo {author} {\bibfnamefont {N.}~\bibnamefont {Killoran}}, \bibinfo {author} {\bibfnamefont {M.}~\bibnamefont {Cramer}}, \ and\ \bibinfo {author} {\bibfnamefont {M.~B.}\ \bibnamefont {Plenio}},\ }\href {\doibase 10.1103/PhysRevLett.112.150501} {\bibfield  {journal} {\bibinfo  {journal} {Phys. Rev. Lett.}\ }\textbf {\bibinfo {volume} {112}},\ \bibinfo {pages} {150501} (\bibinfo {year} {2014})}\BibitemShut {NoStop}%
\bibitem [{\citenamefont {Bouvrie}\ \emph {et~al.}(2017)\citenamefont {Bouvrie}, \citenamefont {Valdés-Hernández}, \citenamefont {Majtey}, \citenamefont {Zander},\ and\ \citenamefont {Plastino}}]{Bouvrie17a}%
  \BibitemOpen
  \bibfield  {author} {\bibinfo {author} {\bibfnamefont {P.}~\bibnamefont {Bouvrie}}, \bibinfo {author} {\bibfnamefont {A.}~\bibnamefont {Valdés-Hernández}}, \bibinfo {author} {\bibfnamefont {A.}~\bibnamefont {Majtey}}, \bibinfo {author} {\bibfnamefont {C.}~\bibnamefont {Zander}}, \ and\ \bibinfo {author} {\bibfnamefont {A.}~\bibnamefont {Plastino}},\ }\href {\doibase https://doi.org/10.1016/j.aop.2017.06.004} {\bibfield  {journal} {\bibinfo  {journal} {Annals of Physics}\ }\textbf {\bibinfo {volume} {383}},\ \bibinfo {pages} {401} (\bibinfo {year} {2017})}\BibitemShut {NoStop}%
\bibitem [{\citenamefont {Lo~Franco}\ and\ \citenamefont {Compagno}(2018)}]{Lo18a}%
  \BibitemOpen
  \bibfield  {author} {\bibinfo {author} {\bibfnamefont {R.}~\bibnamefont {Lo~Franco}}\ and\ \bibinfo {author} {\bibfnamefont {G.}~\bibnamefont {Compagno}},\ }\href {\doibase 10.1103/PhysRevLett.120.240403} {\bibfield  {journal} {\bibinfo  {journal} {Phys. Rev. Lett.}\ }\textbf {\bibinfo {volume} {120}},\ \bibinfo {pages} {240403} (\bibinfo {year} {2018})}\BibitemShut {NoStop}%
\bibitem [{\citenamefont {Mahdavipour}\ \emph {et~al.}(2024)\citenamefont {Mahdavipour}, \citenamefont {Nosrati}, \citenamefont {Sciara}, \citenamefont {Morandotti},\ and\ \citenamefont {Lo~Franco}}]{Mahdavipour24a}%
  \BibitemOpen
  \bibfield  {author} {\bibinfo {author} {\bibfnamefont {K.}~\bibnamefont {Mahdavipour}}, \bibinfo {author} {\bibfnamefont {F.}~\bibnamefont {Nosrati}}, \bibinfo {author} {\bibfnamefont {S.}~\bibnamefont {Sciara}}, \bibinfo {author} {\bibfnamefont {R.}~\bibnamefont {Morandotti}}, \ and\ \bibinfo {author} {\bibfnamefont {R.}~\bibnamefont {Lo~Franco}},\ }\href {\doibase 10.1103/PRXQuantum.5.040350} {\bibfield  {journal} {\bibinfo  {journal} {PRX Quantum}\ }\textbf {\bibinfo {volume} {5}},\ \bibinfo {pages} {040350} (\bibinfo {year} {2024})}\BibitemShut {NoStop}%
\bibitem [{\citenamefont {Piccolini}\ \emph {et~al.}(2024)\citenamefont {Piccolini}, \citenamefont {Karczewski}, \citenamefont {Winter},\ and\ \citenamefont {Lo~Franco}}]{Piccolini_2025}%
  \BibitemOpen
  \bibfield  {author} {\bibinfo {author} {\bibfnamefont {M.}~\bibnamefont {Piccolini}}, \bibinfo {author} {\bibfnamefont {M.}~\bibnamefont {Karczewski}}, \bibinfo {author} {\bibfnamefont {A.}~\bibnamefont {Winter}}, \ and\ \bibinfo {author} {\bibfnamefont {R.}~\bibnamefont {Lo~Franco}},\ }\href {\doibase 10.1088/2058-9565/ad8214} {\bibfield  {journal} {\bibinfo  {journal} {Quantum Science and Technology}\ }\textbf {\bibinfo {volume} {10}},\ \bibinfo {pages} {015013} (\bibinfo {year} {2024})}\BibitemShut {NoStop}%
\bibitem [{\citenamefont {Troiani}\ \emph {et~al.}(2025)\citenamefont {Troiani}, \citenamefont {Angeli}, \citenamefont {Secchi},\ and\ \citenamefont {Pittalis}}]{Troiani25a}%
  \BibitemOpen
  \bibfield  {author} {\bibinfo {author} {\bibfnamefont {F.}~\bibnamefont {Troiani}}, \bibinfo {author} {\bibfnamefont {C.}~\bibnamefont {Angeli}}, \bibinfo {author} {\bibfnamefont {A.}~\bibnamefont {Secchi}}, \ and\ \bibinfo {author} {\bibfnamefont {S.}~\bibnamefont {Pittalis}},\ }\href {\doibase 10.1103/PhysRevB.111.L161110} {\bibfield  {journal} {\bibinfo  {journal} {Phys. Rev. B}\ }\textbf {\bibinfo {volume} {111}},\ \bibinfo {pages} {L161110} (\bibinfo {year} {2025})}\BibitemShut {NoStop}%
\bibitem [{\citenamefont {Vedral}(2003)}]{Vedral03a}%
  \BibitemOpen
  \bibfield  {author} {\bibinfo {author} {\bibfnamefont {V.}~\bibnamefont {Vedral}},\ }\href {\doibase doi:10.2478/BF02476298} {\bibfield  {journal} {\bibinfo  {journal} {Open Physics}\ }\textbf {\bibinfo {volume} {1}},\ \bibinfo {pages} {289} (\bibinfo {year} {2003})}\BibitemShut {NoStop}%
\bibitem [{\citenamefont {Shi}(2004)}]{Shi04a}%
  \BibitemOpen
  \bibfield  {author} {\bibinfo {author} {\bibfnamefont {Y.}~\bibnamefont {Shi}},\ }\href {\doibase 10.1088/0305-4470/37/26/014} {\bibfield  {journal} {\bibinfo  {journal} {Journal of Physics A: Mathematical and General}\ }\textbf {\bibinfo {volume} {37}},\ \bibinfo {pages} {6807} (\bibinfo {year} {2004})}\BibitemShut {NoStop}%
\bibitem [{\citenamefont {Pittalis}\ \emph {et~al.}(2015)\citenamefont {Pittalis}, \citenamefont {Troiani}, \citenamefont {Rozzi},\ and\ \citenamefont {Vignale}}]{Pittalis15a}%
  \BibitemOpen
  \bibfield  {author} {\bibinfo {author} {\bibfnamefont {S.}~\bibnamefont {Pittalis}}, \bibinfo {author} {\bibfnamefont {F.}~\bibnamefont {Troiani}}, \bibinfo {author} {\bibfnamefont {C.~A.}\ \bibnamefont {Rozzi}}, \ and\ \bibinfo {author} {\bibfnamefont {G.}~\bibnamefont {Vignale}},\ }\href {\doibase 10.1103/PhysRevB.91.075109} {\bibfield  {journal} {\bibinfo  {journal} {Phys. Rev. B}\ }\textbf {\bibinfo {volume} {91}},\ \bibinfo {pages} {075109} (\bibinfo {year} {2015})}\BibitemShut {NoStop}%
\bibitem [{\citenamefont {Krylov}(2020)}]{Krylov20a}%
  \BibitemOpen
  \bibfield  {author} {\bibinfo {author} {\bibfnamefont {A.~I.}\ \bibnamefont {Krylov}},\ }\href {\doibase 10.1063/5.0018597} {\bibfield  {journal} {\bibinfo  {journal} {The Journal of Chemical Physics}\ }\textbf {\bibinfo {volume} {153}},\ \bibinfo {pages} {080901} (\bibinfo {year} {2020})}\BibitemShut {NoStop}%
\bibitem [{\citenamefont {Gühne}\ and\ \citenamefont {Tóth}(2009)}]{Guhne09a}%
  \BibitemOpen
  \bibfield  {author} {\bibinfo {author} {\bibfnamefont {O.}~\bibnamefont {Gühne}}\ and\ \bibinfo {author} {\bibfnamefont {G.}~\bibnamefont {Tóth}},\ }\href {\doibase https://doi.org/10.1016/j.physrep.2009.02.004} {\bibfield  {journal} {\bibinfo  {journal} {Physics Reports}\ }\textbf {\bibinfo {volume} {474}},\ \bibinfo {pages} {1} (\bibinfo {year} {2009})}\BibitemShut {NoStop}%
\bibitem [{\citenamefont {T\'oth}\ \emph {et~al.}(2009)\citenamefont {T\'oth}, \citenamefont {Knapp}, \citenamefont {G\"uhne},\ and\ \citenamefont {Briegel}}]{Toth09a}%
  \BibitemOpen
  \bibfield  {author} {\bibinfo {author} {\bibfnamefont {G.}~\bibnamefont {T\'oth}}, \bibinfo {author} {\bibfnamefont {C.}~\bibnamefont {Knapp}}, \bibinfo {author} {\bibfnamefont {O.}~\bibnamefont {G\"uhne}}, \ and\ \bibinfo {author} {\bibfnamefont {H.~J.}\ \bibnamefont {Briegel}},\ }\href {\doibase 10.1103/PhysRevA.79.042334} {\bibfield  {journal} {\bibinfo  {journal} {Phys. Rev. A}\ }\textbf {\bibinfo {volume} {79}},\ \bibinfo {pages} {042334} (\bibinfo {year} {2009})}\BibitemShut {NoStop}%
\bibitem [{\citenamefont {Atkins}\ and\ \citenamefont {Friedman}(2011)}]{Atkins}%
  \BibitemOpen
  \bibfield  {author} {\bibinfo {author} {\bibfnamefont {P.}~\bibnamefont {Atkins}}\ and\ \bibinfo {author} {\bibfnamefont {R.}~\bibnamefont {Friedman}},\ }\href@noop {} {\emph {\bibinfo {title} {Molecular Quantum Mechanics}}}\ (\bibinfo  {publisher} {Oxford University Press},\ \bibinfo {year} {2011})\BibitemShut {NoStop}%
\bibitem [{\citenamefont {Martin}(2020)}]{Martin}%
  \BibitemOpen
  \bibfield  {author} {\bibinfo {author} {\bibfnamefont {R.}~\bibnamefont {Martin}},\ }\href@noop {} {\emph {\bibinfo {title} {Electronic Structure: Basic Theory and Practical Methods}}}\ (\bibinfo  {publisher} {Cambridge University Press},\ \bibinfo {year} {2020})\BibitemShut {NoStop}%
\bibitem [{\citenamefont {Friis}\ \emph {et~al.}(2013)\citenamefont {Friis}, \citenamefont {Lee},\ and\ \citenamefont {Bruschi}}]{Friis13a}%
  \BibitemOpen
  \bibfield  {author} {\bibinfo {author} {\bibfnamefont {N.}~\bibnamefont {Friis}}, \bibinfo {author} {\bibfnamefont {A.~R.}\ \bibnamefont {Lee}}, \ and\ \bibinfo {author} {\bibfnamefont {D.~E.}\ \bibnamefont {Bruschi}},\ }\href {\doibase 10.1103/PhysRevA.87.022338} {\bibfield  {journal} {\bibinfo  {journal} {Phys. Rev. A}\ }\textbf {\bibinfo {volume} {87}},\ \bibinfo {pages} {022338} (\bibinfo {year} {2013})}\BibitemShut {NoStop}%
\bibitem [{\citenamefont {Szalay}\ \emph {et~al.}(2021)\citenamefont {Szalay}, \citenamefont {Zimborás}, \citenamefont {Máté}, \citenamefont {Barcza}, \citenamefont {Schilling},\ and\ \citenamefont {Örs Legeza}}]{Szalay_2021}%
  \BibitemOpen
  \bibfield  {author} {\bibinfo {author} {\bibfnamefont {S.}~\bibnamefont {Szalay}}, \bibinfo {author} {\bibfnamefont {Z.}~\bibnamefont {Zimborás}}, \bibinfo {author} {\bibfnamefont {M.}~\bibnamefont {Máté}}, \bibinfo {author} {\bibfnamefont {G.}~\bibnamefont {Barcza}}, \bibinfo {author} {\bibfnamefont {C.}~\bibnamefont {Schilling}}, \ and\ \bibinfo {author} {\bibnamefont {Örs Legeza}},\ }\href {\doibase 10.1088/1751-8121/ac0646} {\bibfield  {journal} {\bibinfo  {journal} {Journal of Physics A: Mathematical and Theoretical}\ }\textbf {\bibinfo {volume} {54}},\ \bibinfo {pages} {393001} (\bibinfo {year} {2021})}\BibitemShut {NoStop}%
\bibitem [{\citenamefont {Tsukerblat}(2006)}]{Tsukerblat}%
  \BibitemOpen
  \bibfield  {author} {\bibinfo {author} {\bibfnamefont {B.}~\bibnamefont {Tsukerblat}},\ }\href {https://books.google.it/books?id=vmuTAwAAQBAJ} {\emph {\bibinfo {title} {Group Theory in Chemistry and Spectroscopy: A Simple Guide to Advanced Usage}}},\ Dover Books on Chemistry\ (\bibinfo  {publisher} {Dover Publications},\ \bibinfo {year} {2006})\BibitemShut {NoStop}%
\bibitem [{\citenamefont {Karbach}\ \emph {et~al.}(1993)\citenamefont {Karbach}, \citenamefont {M\"utter}, \citenamefont {Ueberholz},\ and\ \citenamefont {Kr\"oger}}]{Karbach93a}%
  \BibitemOpen
  \bibfield  {author} {\bibinfo {author} {\bibfnamefont {M.}~\bibnamefont {Karbach}}, \bibinfo {author} {\bibfnamefont {K.-H.}\ \bibnamefont {M\"utter}}, \bibinfo {author} {\bibfnamefont {P.}~\bibnamefont {Ueberholz}}, \ and\ \bibinfo {author} {\bibfnamefont {H.}~\bibnamefont {Kr\"oger}},\ }\href {\doibase 10.1103/PhysRevB.48.13666} {\bibfield  {journal} {\bibinfo  {journal} {Phys. Rev. B}\ }\textbf {\bibinfo {volume} {48}},\ \bibinfo {pages} {13666} (\bibinfo {year} {1993})}\BibitemShut {NoStop}%
\bibitem [{\citenamefont {Lunkes}\ \emph {et~al.}(2005)\citenamefont {Lunkes}, \citenamefont {Brukner},\ and\ \citenamefont {Vedral}}]{Lunkes05a}%
  \BibitemOpen
  \bibfield  {author} {\bibinfo {author} {\bibfnamefont {C.}~\bibnamefont {Lunkes}}, \bibinfo {author} {\bibfnamefont {C.}~\bibnamefont {Brukner}}, \ and\ \bibinfo {author} {\bibfnamefont {V.}~\bibnamefont {Vedral}},\ }\href {\doibase 10.1103/PhysRevLett.95.030503} {\bibfield  {journal} {\bibinfo  {journal} {Phys. Rev. Lett.}\ }\textbf {\bibinfo {volume} {95}},\ \bibinfo {pages} {030503} (\bibinfo {year} {2005})}\BibitemShut {NoStop}%
\bibitem [{\citenamefont {V\'ertesi}(2007)}]{Vertesi07a}%
  \BibitemOpen
  \bibfield  {author} {\bibinfo {author} {\bibfnamefont {T.}~\bibnamefont {V\'ertesi}},\ }\href {\doibase 10.1103/PhysRevA.75.042330} {\bibfield  {journal} {\bibinfo  {journal} {Phys. Rev. A}\ }\textbf {\bibinfo {volume} {75}},\ \bibinfo {pages} {042330} (\bibinfo {year} {2007})}\BibitemShut {NoStop}%
\bibitem [{\citenamefont {Troiani}(2011)}]{Troiani11a}%
  \BibitemOpen
  \bibfield  {author} {\bibinfo {author} {\bibfnamefont {F.}~\bibnamefont {Troiani}},\ }\href {\doibase 10.1103/PhysRevA.83.022324} {\bibfield  {journal} {\bibinfo  {journal} {Phys. Rev. A}\ }\textbf {\bibinfo {volume} {83}},\ \bibinfo {pages} {022324} (\bibinfo {year} {2011})}\BibitemShut {NoStop}%
\bibitem [{\citenamefont {Siloi}\ and\ \citenamefont {Troiani}(2014)}]{Siloi14a}%
  \BibitemOpen
  \bibfield  {author} {\bibinfo {author} {\bibfnamefont {I.}~\bibnamefont {Siloi}}\ and\ \bibinfo {author} {\bibfnamefont {F.}~\bibnamefont {Troiani}},\ }\href {\doibase 10.1103/PhysRevA.90.042328} {\bibfield  {journal} {\bibinfo  {journal} {Phys. Rev. A}\ }\textbf {\bibinfo {volume} {90}},\ \bibinfo {pages} {042328} (\bibinfo {year} {2014})}\BibitemShut {NoStop}%
\bibitem [{\citenamefont {Breuer}(2005)}]{Breuer05a}%
  \BibitemOpen
  \bibfield  {author} {\bibinfo {author} {\bibfnamefont {H.-P.}\ \bibnamefont {Breuer}},\ }\href {\doibase 10.1088/0305-4470/38/41/013} {\bibfield  {journal} {\bibinfo  {journal} {Journal of Physics A: Mathematical and General}\ }\textbf {\bibinfo {volume} {38}},\ \bibinfo {pages} {9019} (\bibinfo {year} {2005})}\BibitemShut {NoStop}%
\end{thebibliography}


\end{document}